\documentclass[final]{elsarticle}

\usepackage{lineno,hyperref}
\modulolinenumbers[5]

\usepackage{hyperref}
\usepackage[utf8]{inputenc}
\usepackage[cmex10]{amsmath}
\usepackage{amssymb}
\usepackage{graphicx}
\usepackage{mathtools}
\usepackage{caption}
\usepackage{subcaption}
\usepackage{eurosym}
\usepackage{algorithm}
\usepackage{algpseudocode}
\usepackage[noadjust]{cite}
\usepackage{listings}
\usepackage{color}
\usepackage{multirow}
\usepackage[]{units}
\usepackage{xspace}
\usepackage{tikz}

\newtheorem{thm}{Theorem}
\newtheorem{lem}[thm]{Lemma}
\newproof{pf}{Proof}

\newcommand{\cindependent}{\emph{c-independent}\xspace}
\newcommand{\dindependent}{\emph{d-independent}\xspace}
\newcommand{\cworkload}{\emph{c-workload}\xspace}
\newcommand{\dworkload}{\emph{d-workload}\xspace}
\newcommand{\chungarian}{\emph{c-hungarian}\xspace}
\newcommand{\cssi}{\emph{c-greedy}\xspace}

\begin{document}

\begin{frontmatter}

\title{Decentralized dynamic task allocation for UAVs with limited communication range}

\author[mymainaddress]{Marc~Pujol-Gonzalez}
\ead{mpujol@iiia.csic.es}

\author[mymainaddress]{Jesus~Cerquides}
\ead{cerquide@iiia.csic.es}

\author[mymainaddress]{Pedro~Meseguer}
\ead{pedro@iiia.csic.es}

\author[mymainaddress]{Juan~A.~Rodriguez-Aguilar\corref{mycorrespondingauthor}}
\cortext[mycorrespondingauthor]{Corresponding author}
\ead{jar@iiia.csic.es}

\author[mysecondaryaddress]{Milind Tambe}
\ead{tambe@usc.edu}

\address[mymainaddress]{Artificial Intelligence Research Institute of the Spanish National Research Council (IIIA-CSIC), Campus de la UAB, 08193 Bellaterra, Spain}
\address[mysecondaryaddress]{University of Southern California, Los Angeles, CA 90089}

\begin{abstract}
We present the Limited-range Online Routing Problem (LORP), which involves a team of Unmanned Aerial Vehicles (UAVs) with limited communication range that must autonomously coordinate to service task requests. We first show a general approach to cast this dynamic problem as a sequence of decentralized task allocation problems. Then we present two solutions both based on modeling the allocation task as a Markov Random Field to subsequently assess decisions by means of the decentralized Max-Sum algorithm. Our first solution assumes independence between requests, whereas our second solution also considers the UAVs’ workloads. A thorough empirical evaluation shows that our workload-based solution consistently outperforms current state-of-the-art methods in a wide range of scenarios, lowering the average service time up to 16\%. In the best-case scenario there is no gap between our decentralized solution and centralized techniques. In the worst-case scenario we manage to reduce by 25\% the gap between current decentralized and centralized techniques. Thus, our solution becomes the method of choice for our problem.
\end{abstract}


\end{frontmatter}



\section{Introduction}
\label{sec:introduction}

Unmanned Aerial Vehicles (UAVs) are 
an attractive technology for large-area surveillance~\citep{kingston2008decentralized}. 
Today, there are readily available UAVs that are reasonably cheap, have many sensing abilities, exhibit a long endurance and can communicate using radios. 
Several applications can be efficiently tackled with a team of such UAVs:
power line monitoring, fire detection, and disaster response among others~\citep{civiluavassesment}.

UAVs have traditionally been controlled either remotely or by following externally-designed flight plans. Requiring human operators for each UAV implies a large, specialized and expensive human workforce. Likewise, letting UAVs follow externally prepared plans introduces a single point of failure (the planner) and requires UAVs with expensive (satellite) radios to maintain continuous communication with a central station. These constraints are acceptable in some application domains, other applications require more flexible techniques.

For instance, consider a force of park rangers tasked with the surveillance of a large natural park. Upon reception of an emergency notification, the rangers must assess the situation as quickly as possible. With this aim, they could deploy a team of UAVs continuously fly throughout the park. Thereafter, they could issue requests for their UAVs to check certain locations. To maintain the cost-effectiveness of the approach, such UAVs cannot employ expensive communication devices. Thus, the UAVs would have limited communication ranges, oftentimes significantly smaller than the park's extension. Notice that, in this setting, neither human remote control nor centralized planning is feasible due to such communication constraints.

In this scenario, a possibility would be to deploy the planes with a fixed mission, so that they just go to the desired location(s), perform any required checks and fly back to some base. However, this approach is grossly unequipped to handle the dynamism of the problem. That is, during a critical situation (e.g., when there is an actual fire in the park), the rangers will receive many reports, and will quickly run out of UAVs to handle them. A possibly better approach involves UAVs that can act autonomously and can coordinate between them. Such UAVs keep flying around the park and coordinate to attend requests as they are introduced into the system by rangers, thus hopefully improving their effectiveness.

Nevertheless, the autonomous operation and coordination of UAVs is an open research question receiving increasing attention. It involves challenges ranging from low-level operational details of flight control to high level coordination between UAVs~\citep{national2006Decadal}. In this paper we assume that low level control can be handled entirely by the UAVs' auto-pilot systems, and focus on high-level coordination challenges instead.


There are two main approaches to enable such coordination in the literature, each best suited for a different kind of UAV missions. On the one hand, the objective in exploratory missions is to collect accurate information and keep it updated. Hence, the first approach focuses on shared information collection, fusion and maintenance techniques to fulfill these missions~\citep{zlot2002multi}. 
On the other hand, some applications involve specific tasks that the UAVs should carry. For instance, check requests introduced by the rangers can be considered as tasks to be performed by the UAVs. Hence, coordination between UAVs in this context implies making decisions about which UAV should conduct each task. 

Unfortunately, most state-of-the-art multi-agent allocation mechanisms cannot be employed in our settings, particularly due to the communication range limitation. Hence, we first identify the characteristics and particular challenges of such applications and define the Limited-range Online Routing Problem (LORP) to capture them. After studying these characteristics, we advocate for a solution approach where UAVs make quick decisions based solely on local information, where the neighbors of a UAV are those UAVs with which it can directly communicate at a given point in time.

Next, we introduce a solution approach based on representing the problem as a factorized utility function (actually a Markov Random Field) and optimizing it using the Max-Sum distributed algorithm. This technique provides two significant advantages: (i) the problem is represented as a single global utility function. Hence, the collective behavior of the UAVs is easier to predict and reason about than when it emerges from the definition of individual behaviors; and (ii) we can draw results from the extensive Max-Sum literature, including its theoretical properties (e.g.: convergence and quality guarantees~\citep{weiss2001optimality,salgado2011exploiting}) and encouraging experimental results~\citep{rogers2011bounded}.

We present an initial development of this solution in Section~\ref{sec:maxsum}, assuming that the cost of servicing each request is independent of other requests assigned to the UAV. Thereafter, we introduce a second solution where UAVs adjust their estimations of the cost of servicing a task depending on their workload, with a slight increment in complexity. As a result, UAVs using this solution are better equipped to dynamically capture and exploit the distribution of incoming requests.

The contributions of this paper are the following:
\begin{itemize}
\item We introduce the LORP and identify its need for quick loops of assessment, decision and action in highly dynamic, communication constrained UAV applications.
\item We propose to cast the LORP as an optimization problem and distributedly solve it using Max-Sum, thus providing great flexibility to introduce new heuristics without modifying the solving algorithm.
\item We capitalize on this advantage by introducing the workload heuristic, which exploits the dynamic characteristics of the problem to make better decisions.
\item We empirically evaluate the proposed algorithms, showing that: (i) the proposed approach allows for effective request allocation in a highly dynamic, communication constrained domain; and (ii) our workload-based solution achieves between 6\% and 16\% lower service times than current state-of-the-art methods, and its actual performance comes very close to that of centralized solutions.
\end{itemize}

The contributions above build upon and extend the preliminary work presented in~\citep{pujol2013engineering}. However, notice that here we improve that work by: (i) relating it to current state-of-the-art literature, (ii) presenting the LORP as a general problem and showing how to cast it as a distributed task allocation problem, (iii) providing proofs of the results that allow us to improve Max-Sum's efficiency, and (iv) performing a thorough empirical comparison between the proposed algorithms and the related methods.


The rest of the paper is organized as follows. First, Section~\ref{sec:state-of-art} reviews current coordination approaches. Next, Section~\ref{sec:problem-definition} describes our problem in detail, along with an example scenario. Thereafter we present a first MRF-based solution in Section~\ref{sec:maxsum}, explaining how the LORP can be encoded and efficiently solved using Max-Sum. Then, Section~\ref{sec:workload} shows how to improve this representation to better spread the workload between planes. Finally, Sections~\ref{sec:results}~and~\ref{sec:results2} empirically evaluate our approach, and Section~\ref{sec:conclusions} presents some conclusions and further lines of work.

\section{Related Work}
\label{sec:state-of-art}

There exists a large body of research related to multi-agent (and/or multi-robot) task allocation.

Our problem is actually a variant of the Single-Task, Single-Robot (ST-SR) dynamic Multiple Travelling Salesman Problem (MTSP) as described in Gerkey's taxonomy of multi-robot task allocation systems~\citep{gerkey2004formal}. In the MTSP,  a number of robots (or agents) have to visit a number of waypoints in the minimum possible time, while new waypoints may be introduced at any time. In fact, our variant is even harder than the MTSP because of the communication range limitation, and hence it is also strongly NP-Hard. 

Most works on the MSTP focus on market-based allocation mechanisms~\citep{dias2000free,gerkey2002sold,dias2006market,dash2007allocation}, where tasks are allocated to agents using auctions.
The pre-cursor in this area is the auction algorithm~\citep{bertsekas1988auction}, an iterative distributed algorithm to optimally solve the classic \emph{assignment problem}~\citep{kuhn1955hungarian}. The assignment problem is a static task allocation problem where several tasks can be allocated to a number of agents. Each agent has a different cost for performing each task, and can only be assigned to a single task. Then, the objective is to find the allocation that minimizes the total cost incurred after assigning all tasks.

Nonetheless, on most actual-world applications there may be more tasks than agents. If the cost for a robot to perform two tasks is simply the addition of the costs of performing each task separately, then the above algorithm still yields an optimal allocation. However, in routing problems the cost of visiting multiple waypoints depends on the relative positions between those waypoints. Hence, the auction algorithm is not optimal in this case, and can lead to arbitrarily bad solutions~\citep{gerkey2004formal}.

A first step to mitigate this problem is introduced in~\citep{gerkey2002sold}. Instead of performing an iterative auction, the idea here is to perform a parallel single-item auction for each task. These auctions can be resolved in a single round. Afterwards, tasks are continuously re-auctioned, but with agents' bids updated to account for the new state of the world. This approach, described as the Parallel Single-Item (PSI) auctions mechanism in~\citep{koenig2010progress}, improves the robustness of the system and equips it to handle changing conditions and/or the introduction of new tasks.
%
Although PSI auctions capture some of those dependencies by reallocating tasks when conditions change, another possibility is to explicitly consider them. In the Sequential Single-Item auctions~\citep{koenig2006power} method,  agents perform an iterative auction where a single task is assigned at each iteration. After each round, agents update their bids according to the tasks they already got in previous iterations, hence taking into account the synergies between tasks. This approach even has quality guarantees in static domains~\citep{lagoudakis2005auction}, but is not well-suited for dynamic domains because it requires both an arbitrary number of communication cycles and global communication.

An entirely different approach to task allocation is to employ consensus algorithms~\citep{alighanbari2005decentralized,li2010consensus} to let agents maintain consistent information about the system's state. After reaching consensus, each agent can independently compute its plan using a deterministic algorithm, and the resulting plans are guaranteed to be consistent with each other. The main advantage of these algorithms is that they have been shown to converge even on time-varying network topologies. However, the time required to reach such convergence is unbounded and they cannot converge during network partitions. Hence, consensus-based works typically strive to either guarantee network connectedness~\citep{mosteo2008multi} or minimize the impact of such a potentially large decision making time. For instance, \citep{uavfiremon} presents a UAV coordination model for fire perimeter tracking, where UAVs reach a consensus on which section of the perimeter is to be monitored by each UAV. When the conditions change, the UAVs keep monitoring their section until a new consensus is eventually reached. Because the fire perimeter is unlikely to evolve very rapidly, the previous assignment works relatively well even if takes a long time to reach a new consensus.

Finally, some researchers tried to combine the advantages of market-based mechanisms (quick decisions) with those of the consensus-based ones (eventual consistency)~\citep{zhao2016allocation}. Along these lines, Choi~et.~al.~\citep{choi2009consensus} introduced the Consensus Based Bundle Algorithm (CBBA), that directly reaches a consensus on (an auction-based) solution. Nonetheless, this approach still suffers from the low resiliency to network disconnections.

As detailed in the next section, LORP scenarios are highly dynamic and involve very rapid changes in the communication network topology, including frequent disconnections. Hence, consensus-based approaches do not seem particularly well suited for this domain. Moreover, we can only employ the simpler auction-based approaches (such as PSI) due to the lack of global communication.


\section{The limited-range online routing problem}
\label{sec:problem-definition}

In this section we present the Limited-range Online Routing Problem. We first describe the problem intuitively and present an example scenario. Thereafter, we show how it can be cast as a distributed task allocation problem.

\subsection{Motivating Example} 

\begin{figure}
    \centering
	\includegraphics[width=0.6\linewidth]{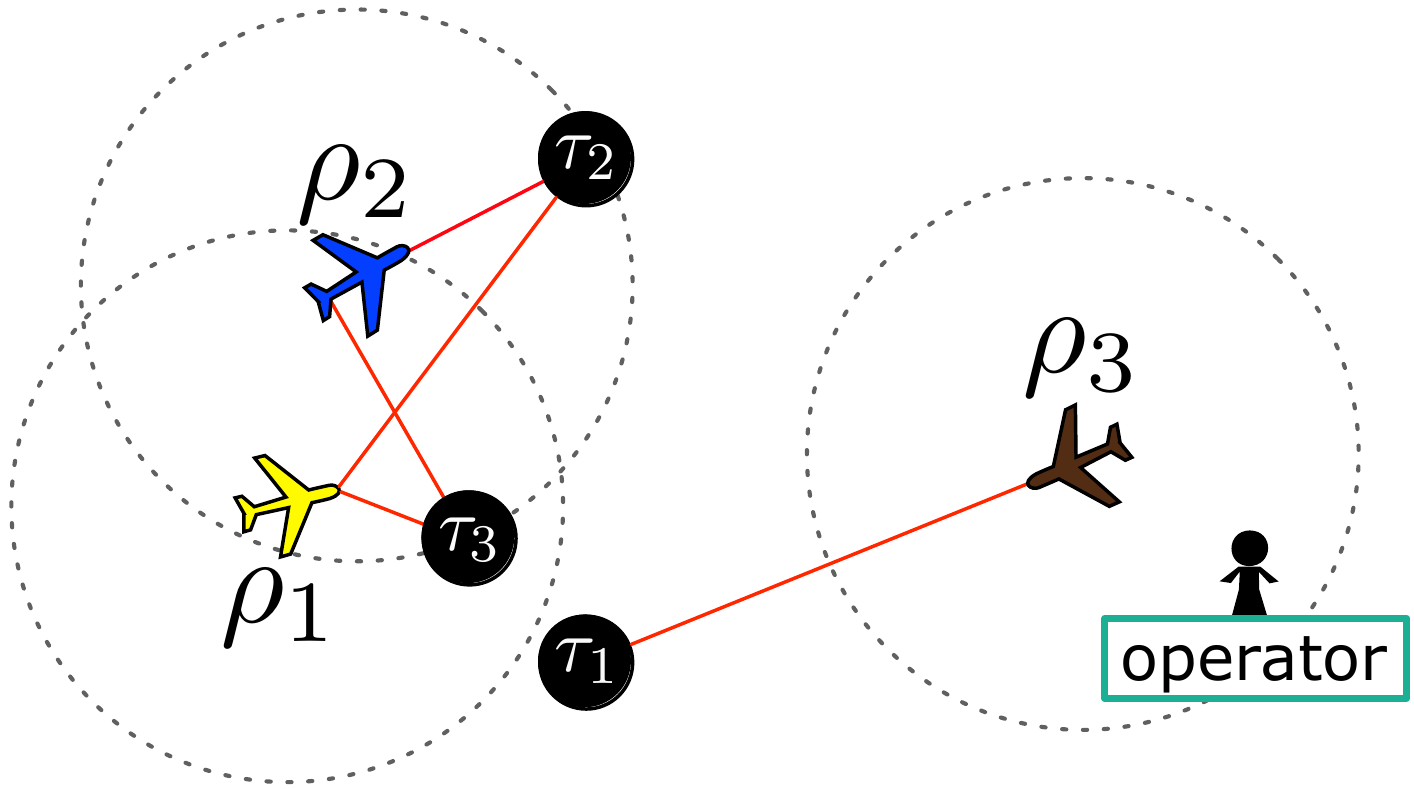}
	\caption{\label{fig:scenario} Example LORP scenario.}
\end{figure}


Consider a force of park rangers that acquires a fleet of unmanned UAVs to help them monitor a large natural park. The force intends to request its UAVs to check certain locations when alerts are received, or as part of their routine surveillance plans. In turn, UAVs are expected to fulfill these requests as quickly as possible. Specifically, UAVs should try to minimize the average request service time, where service time is the amount of time passed between the issue of a request by an operator and the arrival of an UAV to the request's location.\footnote{We assume that a request is serviced as soon as a UAV reaches its location.} 
A common approach to this kind of problems is to adopt a centralized strategy: UAVs' routes are planned at a central station, 
which is in charge of commanding them. Therefore, the central station guarantees cooperation between UAVs. 
However, if the natural park is significantly larger than the UAVs communication ranges, performing centralized planning is unfeasible because the resulting plan can not be effectively transmitted to UAVs.
Figure~\ref{fig:scenario} represents a snapshot of what the rangers' scenario may look like at some point in time. In this example, the force owns three UAVs labeled $\rho_1$, $\rho_2$, and $\rho_3$. The communication range of each UAV is drawn as a dotted circle around it. By inspecting these circles, we observe that UAVs $\rho_1$ and $\rho_2$ can communicate between them. Moreover UAV $\rho_3$ can communicate with the rangers' operator. However, there is no way for UAVs $\rho_1$ and $\rho_2$ to communicate neither with UAV $\rho_3$ nor with the operator. The scenario contains three targets represented by the black circles labelled with $\tau_1$, $\tau_2$, and $\tau_3$. A solid line linking a UAV with a target indicates that the UAV is aware of that target. For instance, UAV $\rho_1$ and UAV $\rho_2$ know that targets $\tau_2$ and $\tau_3$ exist, but they are not aware of target $\tau_1$. Given this scenario, a centralized planner would probably send UAV $\rho_1$ to targets $\tau_3$ and $\tau_1$, UAV $\rho_2$ to target $\tau_2$, and leave $\rho_3$ idle. However, this plan of action can never be computed nor implemented when assuming that UAVs have limited communication range, because there is simply no way for $\rho_1$ to discover $\tau_1$ at this particular point in time.

Notice that the LORP itself does not specify how agents should represent and relay information about which requests are pending and/or completed, nor which messages can be exchanged. Hence, it is part of a LORP solution to define a specific model and an algorithm that allows agents to make specific decisions.
Moreover, within this decision lies an essential trade-off that designers must face. Intuitively, computing more accurate plans improves the quality of the solution \emph{to the currently known part of the problem}, and hence it is a desirable goal. However, the better plans we want to compute, the more time our UAVs have to invest into it, thus hindering their ability to adapt to the new incoming requests. In contrast, we advocate for an approach specifically focused towards better handling \emph{the unknown part of the problem}, as explained in the following section.

\subsection{Approach}
\label{sec:approach}
Notice that, unlike our example scenario above, the LORP is a highly dynamic problem where UAVs constantly move and new requests can be introduced at any time. As a result, we argue that LORP solutions should strive to make quick decisions. On the one hand, a long decision making process can lead to arbitrarily bad decisions being made. That is, a decision that was good when the decision making process started may end up being arbitrarily bad if the process itself took too much time (during which the scenario's conditions kept evolving). On the other hand, making complex decisions (such as long term plans for each UAV) is futile, because there's a high probability that these plans are invalidated. Likewise, operating in a decentralized manner introduces an additional challenge: how to spread and maintain a consistent view of the system between agents. For instance, a LORP solution must specify how UAVs notify others of newly introduced requests, as well as of already serviced ones.

In our approach, we deal with these issues by defining three main processes that are going on simultaneously:
\begin{enumerate}

\item At any time, an operator may introduce a new request by notifying a single UAV in its range, which becomes the request's \emph{owner}. If no UAV is currently in its range, the operator must wait until one comes around. So long as a UAV is the owner of a request, it is responsible for the eventual servicing of that request. However, ownership of a request may be transferred to another UAV as explained next. Since there is exactly one owner of each request at a particular point in time, this guarantees a consistent (yet distributed) view of the system by the UAVs.

\item Concurrently, UAVs run cycles of a request reallocation process. The reallocation process is a sequential process with three phases. First, the UAVs construct a snapshot of the current situation by broadcasting their location and the requests they currently own. Next, the UAVs run some decentralized request allocation algorithm based on the collected information. This phase may take several communication rounds between neighboring UAVs, but such number of rounds must be strictly limited and known beforehand. At the end of these rounds, the algorithm must have decided on which plane should own each request. Finally, these decisions are executed by exchanging the requests' ownership between planes. When a full reallocation cycle finishes, the UAVs start a new one with updated information.

\item Meanwhile, each UAV flies towards the nearest it owns or tries to get in range of the closest operator if it does not own any request. When a UAV reaches the location of a request it owns, the request is considered serviced and removed from the system.
\end{enumerate}

We argue that this approach is particularly well suited for the LORP problem. First, the ownership concept frees the allocation algorithms from having to maintain a consistent situational awareness because only the current owner of a request may decide to allocate it to another UAV. Additionally, by enforcing short allocation cycles the UAVs can make decisions quickly and re-consider them as the situation evolves. 
Obviously, enforcing quick allocations prevents us from employing sophisticated methods that best consider the currently known part of the problem. However, we expect the benefit of being able to constantly reconsider those decisions to make up for --and even surpass-- the loss in plan accuracy from employing simpler allocation methods.

Arguably, another caveat of this approach is that it gives up the possibility of a more controlled network deployment that tries to maximize the UAV's coverage of space, as introduced in other works~\citep{mosteo2008multi,acevedo2013decentralized}. Nonetheless, our proposal provides several significant qualities that those works cannot achieve. First and foremost, our proposal is inherently robust to failures. Even if a massive disruption renders several UAVs useless, our system will remain operative to make the best possible use of the remaining ones. Second, by avoiding any centralization whatsoever, our approach can scale to arbitrarily large numbers of tasks, UAVs and space to cover. Finally, it allows for the emerge of self-organized network structures adapted to the situation's characteristics. Because this characteristic is hard to understand, next we introduce a small example that illustrates how our approach works, and highlights how the UAVs may self-organize while following it.

\subsection{Example of operation}

\begin{figure}
	\centering
	\begin{subfigure}{.9\linewidth}
		\includegraphics[width=\linewidth]{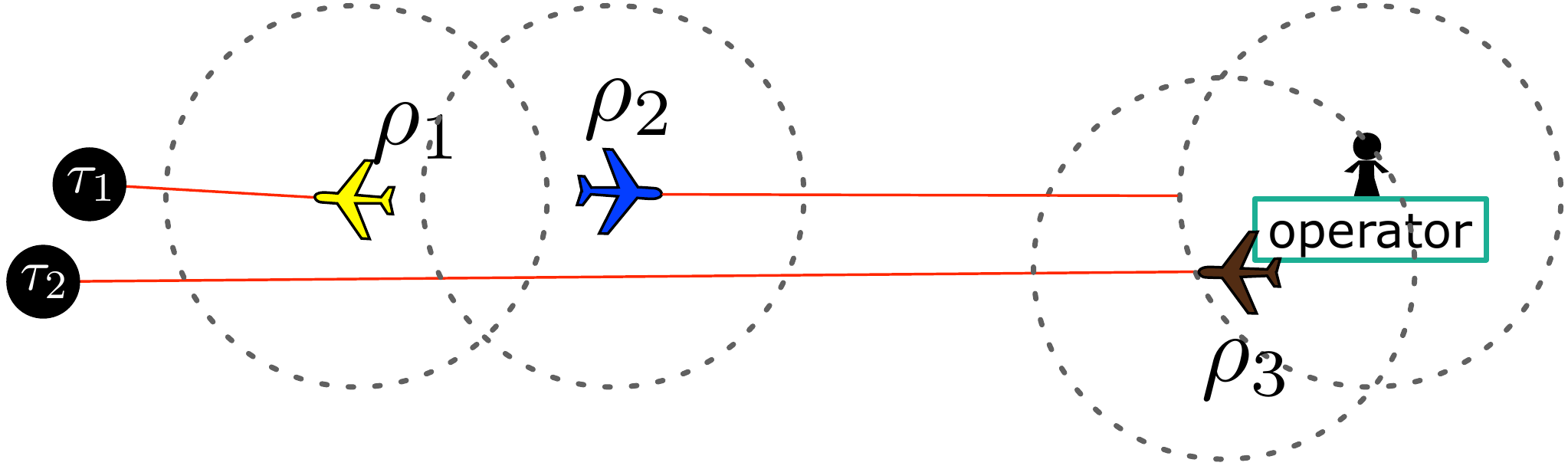}
		\caption{Initial situation \label{fig:example-relay1}}
	\end{subfigure}
	\begin{subfigure}{.9\linewidth}
		\vspace{.5em}
		\includegraphics[width=\linewidth]{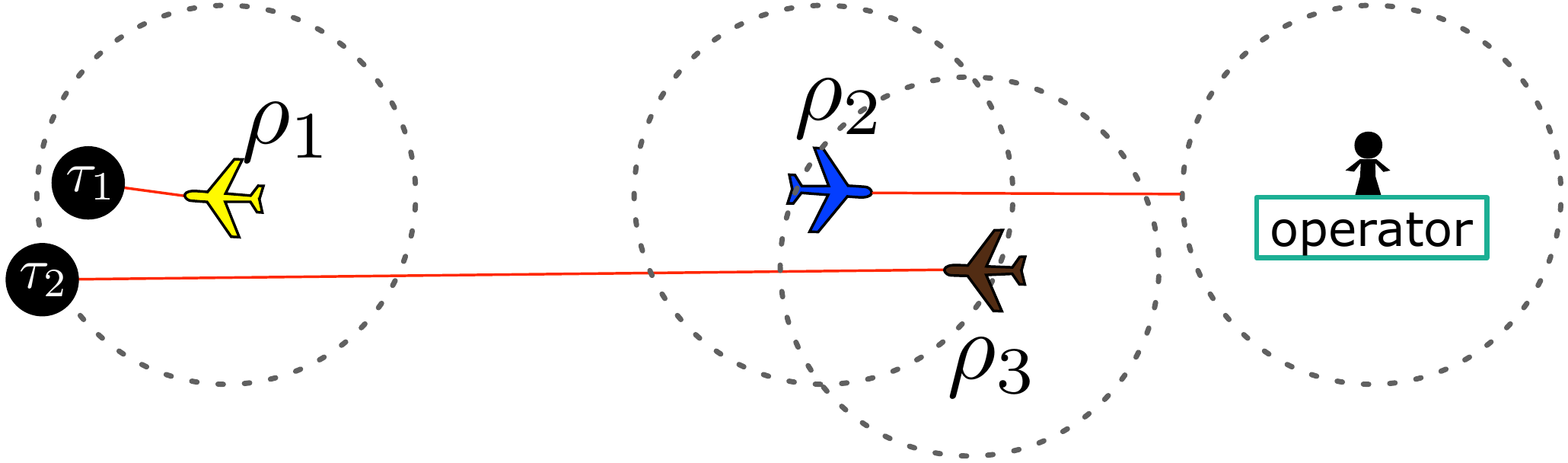}
		\caption{A reallocation negotiation begins \label{fig:example-relay2}}
	\end{subfigure}
	\begin{subfigure}{.9\linewidth}
		\vspace{.5em}
		\includegraphics[width=\linewidth]{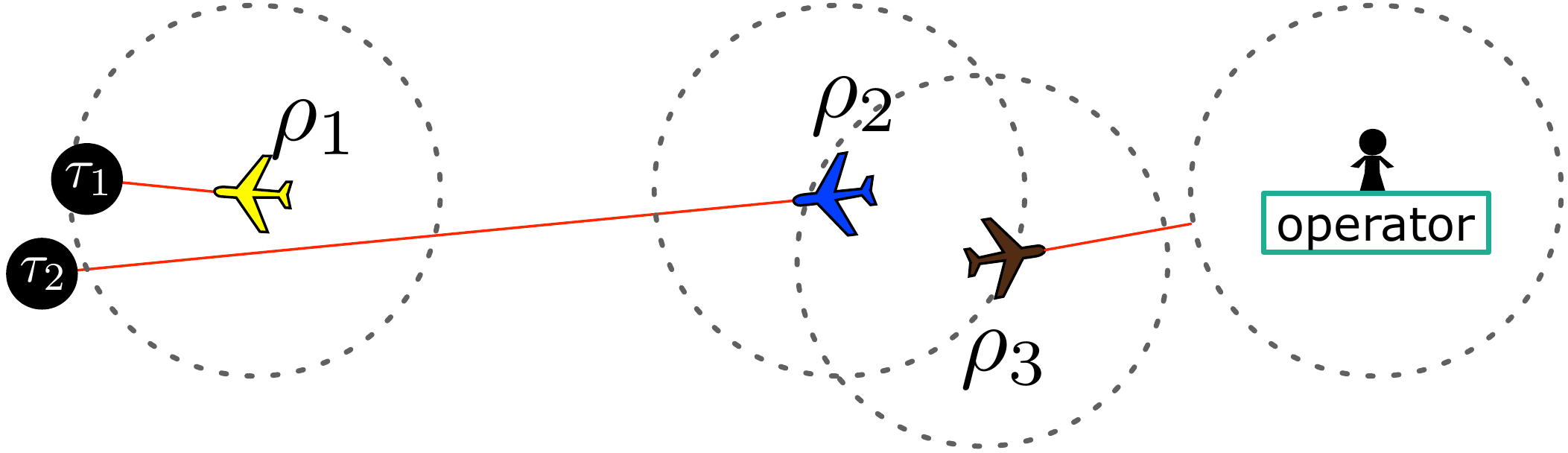}
		\caption{Task relaying is now complete \label{fig:example-relay3}}
	\end{subfigure}
	\caption {Example of task relaying without an enforced communication chain. \label{fig:example-relay}}
\end{figure}

Consider the example in Figure~\ref{fig:example-relay1}. Here UAV $\rho_2$ has no tasks assigned, and hence is flying back to get into operator range (as per process 3 explained above). Meanwhile, UAV $\rho_1$ is trying to complete task $\tau_1$, whereas UAV $\rho_3$ just got task $\tau_2$ from the operator, and starts moving towards its location (also because of process 3). After a while, as shown in Figure~\ref{fig:example-relay2}, UAVs $\rho_2$ and $\rho_3$ get into communication range, and start a reallocation round (process 2). At this point, a sensible allocation method probably assigns task $\tau_2$ to UAV $\rho_2$. Hence, $\tau_2$'s ownership is transferred from UAV $\rho_3$ to UAV $\rho_2$, which becomes the new owner of this task. Finally, (by process 3) $\rho_2$ now advances towards $\tau_2$ whereas $\rho_3$ goes back to the operator to gather new tasks, as shown in Figure~\ref{fig:example-relay3}. Notice that, if further tasks are introduced around $\tau_1$ and $\tau_2$, UAVs $\rho_2$ and $\rho_3$ will keep acting as ``messengers'' bringing those tasks from the operator to UAV $\rho_1$
as illustrated here. Thus, this example shows how UAVs following our approach can naturally form task relay chains, even if the actual tasks are further than what an connected communication chain could reach.

\section{Coordination using Independent Valuations}
\label{sec:maxsum}

The work in~\citep{butterfield2008multi} shows that it is possible to model multi-agent problems as MRFs, and then let agents find the Maximum-A-Posteriori (MAP) of that field to make their decisions. In this section we show how a snapshot of a LORP problem can be encoded as a MRF, assuming that requests are independent between them (i.e,\ that the cost for a UAV to service two requests is the same as the sum of costs of servicing them separately). Thereafter we use Max-Sum to let UAVs compute an allocation in a distributed manner.

\subsection{Problem Encoding as Binary MRF}
\label{subsec:encoding}

Prior to encoding our problem, some notation is in place. Henceforth, 
Let 
 $R = \{\tau_1,\ldots,\tau_m\}$ be a set of requests, $P = \{\rho_1,\ldots,\rho_n\}$ be a set of UAVs, $r$ and $p$ be indexes for requests and UAVs respectively,  $R_p \subseteq R$ 
 be the set of requests that UAV $\rho_p$ can service, and $P_r \subseteq P$ 
 be the set of UAVs that can service request $\tau_r$. A naive encoding 
 of the requests-to-UAVs allocation 
 as an MRF is:
\begin{itemize}
\item Create a variable $x_r$ for each request $\tau_r$. The domain of this variable is the set of UAVs that can service the request, namely $P_r$. If $x_r$ takes value $\rho_p$, it means that request $\tau_r$ will be serviced by UAV $\rho_p$.
\item Create an constraint $c_p$ for each UAV $\rho_p$, that evaluates the cost of servicing the requests assigned to $\rho_p$. 
We assume independence, so the cost of servicing a set of requests is the sum of the costs of servicing each request.
\end{itemize}

$X_p$ is the set of variables that have $\rho_p$ in their domains.
An assignment of values to each of the variables in $X_p$ is noted as $\mathbf{X_p}$.
Solving the MAP problem amounts to finding the combination of request-to-UAV assignments $\mathbf{X^*}$ that satisfies $\mathbf{X^*} = \arg\min_{\mathbf{X}} \sum_{p \in P} c_p(\mathbf{X_p})$.


\begin{figure}[htb]
\begin{subfigure}{.5\linewidth}
\centering
\includegraphics[width=0.75\linewidth]{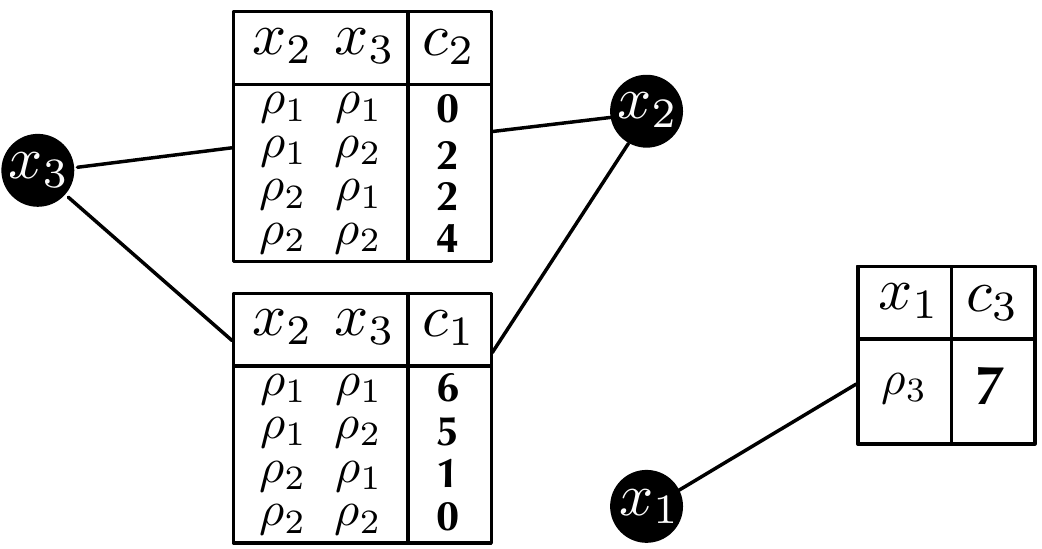}
\caption{Naive MRF encoding.}
\label{fig:ms-naive-encoding}
\end{subfigure}%
\begin{subfigure}{.5\linewidth}
\centering
\includegraphics[width=0.75\linewidth]{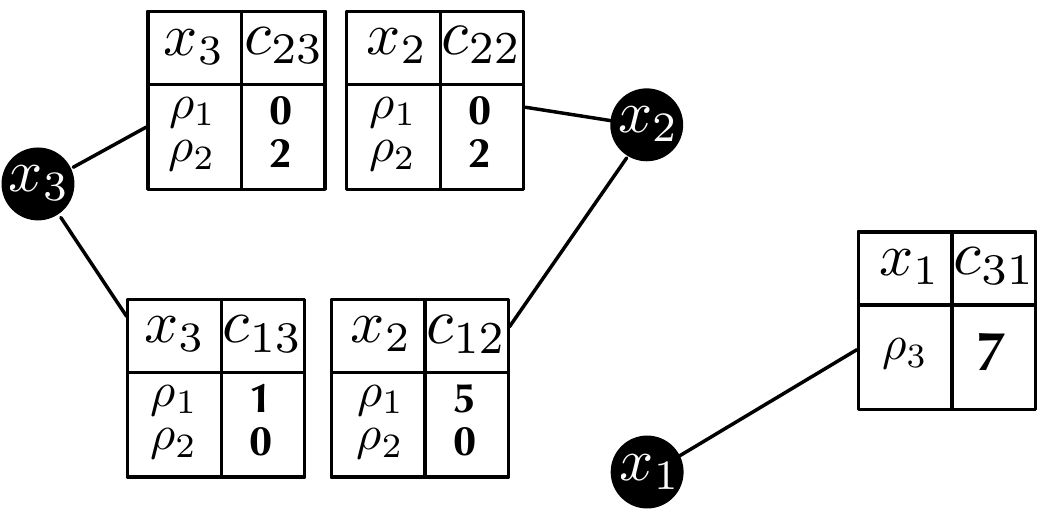}
\caption{Independent valuation.}
\label{fig:ms-independent-tasks}
\end{subfigure}\\[1ex]
\centering
\begin{subfigure}{\linewidth}
\centering
\includegraphics[width=0.75\linewidth]{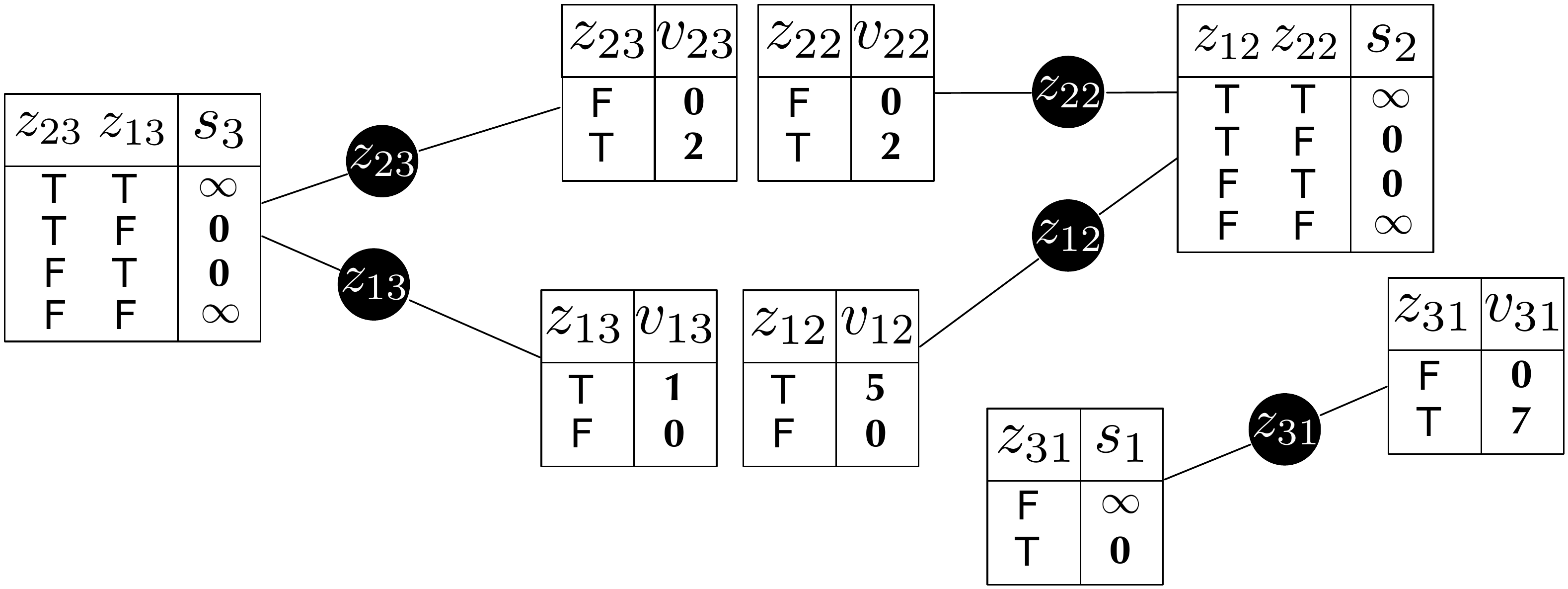}
\caption{\label{fig:ms-independent-binary} Independent task valuation, binary encoding.}
\end{subfigure}
\caption{Encoding the example LORP scenario.}
\end{figure}


Figure~\ref{fig:ms-naive-encoding}  shows an encoding of the motivating example. 
There is a variable for each request.
The domain of $x_1$ is the set of UAVs that can service request $\tau_1$. This is effectively the set of all UAVs that are in communication range of the owner of $\tau_1$. Hence, the domain of $x_1$ is just $\{\rho_3\}$.
Likewise, the domain of $x_2$ and $x_3$ is $\{\rho_1,\rho_2\}$ because both UAVs can fulfill it. 
Next, we create a function $c_p$ for each UAV $\rho_p$. Because $\rho_3$ can only service $\tau_1$, the scope of function $c_3$ is $x_1$. As a result, $c_3$ is a unary function that specifies the cost for UAV $\rho_3$ to service $\tau_1$, namely the distance between $\rho_3$ and $\tau_1$ (hereafter $\delta_{pr}$ will be employed as a shorthand for the distance between $\rho_p$ and $\tau_r$).
The scope of $c_2$ is $\{x_2, x_3\}$, because UAV $\rho_2$ can service both $\tau_2$ and $\tau_3$. Hence, $c_2$ specifies four costs for $\rho_2$:
\begin{enumerate}
\item 0 if both requests are allocated to $\rho_1$.
\item 2 ($\delta_{23}$) if $\tau_2$ is allocated to $\rho_1$ but $\tau_3$ is allocated to $\rho_2$.
\item 2 ($\delta_{22}$) if $\tau_2$ is allocated to $\rho_2$, but $\tau_3$ is allocated to $\rho_1$.
\item 4 ($\delta_{22} + \delta_{23}$) if both requests are allocated to $\rho_2$.
\end{enumerate}
$c_1$ is similarly computed. 
From Figure~\ref{fig:ms-naive-encoding} costs, the optimal assignment is
$\mathbf{X^*} = \langle x_1=\rho_3, x_2=\rho_2, x_3=\rho_1 \rangle$.



This encoding scales poorly. First, 
it does not exploit the fact that we assume independence when computing the cost of servicing a combination of requests in the constraints $c_p$. The number of entries in $c_p$ is the product of the domain sizes of each of the variables in its scope. %
Hence, the number of entries in $c_p$ scales exponentially with respect to the number of requests that UAV $\rho_p$ can service. 
However, we can exploit the independence between requests by decomposing each cost function $c_p$ 
into smaller cost functions, each one evaluating the cost of servicing a single request.
That is, thanks to that independence between requests, we can represent $c_p$ as a combination of cost functions $c_{pr}$, one per variable in the scope of $c_p$, such that $c_p(\mathbf{X_p}) = \sum_{x_r \in X_p}c_{pr}(\mathbf{x_r})$. 
Now, the number of values to specify the cost of servicing a set of requests scales linearly with respect to the number of requests.
Figure~\ref{fig:ms-independent-tasks} represents the example in Figure~\ref{fig:ms-naive-encoding} using this new encoding. Notice that for each UAV we specify the cost of servicing a given request when the request is assigned to it, or 0 
when allocated to another UAV. 
However, the new encoding still suffers from redundancy. 
Say that another UAV $\rho_4$ is in the communication range of both $\rho_1$ and $\rho_2$. Since this UAV would be eligible to serve requests $\tau_2$ and $\tau_3$, the domain of $x_2$ and $x_3$ would become $\{\rho_1,\rho_2,\rho_4\}$. As a result, UAV $\rho_1$ must extend its cost function $c_{12}$ to include a new entry where $\tau_2$ is assigned to $\rho_4$, whose cost is obviously 0. 
%
Therefore, we must aim at an encoding such that a cost function $c_{pr}$ contains only two values: $\delta_{pr}$ if $\tau_r$ is allocated to $\rho_p$, or $0$ otherwise. 



With this aim, we now convert the request variables 
into binary variables, 
replacing each original variable $x_r \in X$ by a set of binary variables $z_{pr}$, one per UAV in $P_r$.
Previous $c_{pr}$ cost functions now generate $v_{pr}$ cost functions on these binary variables.
In addition, for each $r$, $z_{pr}$ are linked through a selection function $s_r$ 
to ensure that a request can be only serviced by a single UAV. For instance, consider variable $x_2$ with domain $\{\rho_1,\rho_2\}$. We create two binary variables $z_{12}$ and $z_{22}$. Intuitively, $z_{12}$ being ``on'' means that request $\tau_2$ is assigned to UAV $\rho_1$. A selection factor linked to both $z_{12}$ and $z_{22}$ would guarantee that only one of the two variables is set to ``on''. In our example, this selection function is a cost function $s_2$, which introduces an infinite cost whenever there is no single variable active. Figure~\ref{fig:ms-independent-binary} shows the binary encoding of the example in Figure~\ref{fig:ms-independent-tasks}.


\subsection{Solving the Problem with Max-Sum}
\label{subsec:solving}

Now we optimize the Max-Sum algorithm to run on the last encoding of Section~\ref{sec:maxsum} (Figure~\ref{fig:ms-independent-binary}). 
Max-Sum sends messages from factors to variables and from variables to factors.
However, our factor graph allows for some simplifications. Notice that each $z_{pr}$ is only linked to cost function $v_{pr}$ and to selector function $s_r$. 
It is direct to observe that the message that $z_{pr}$ must send to $v_{pr}$ is exactly the one received from $s_r$, while the message that it must send to $s_r$ is exactly the  one received from $v_{pr}$. Then, since each variable simply relays messages between the cost function and selection function it is linked to, henceforth we will disregard variables' messages and instead we consider that functions directly exchange messages.    


The Max-Sum general message expression from function $f$ to function $g$ is
\begin{eqnarray}
\resizebox{.92\linewidth}{!}{$\displaystyle
\mu_{f \rightarrow g}(\mathbf{Z_{f \cap g}}) = \min_{\mathbf{Z_{f - g}}} \big[ f(\mathbf{Z_{f-g}},\mathbf{Z_{f \cap g}}) + \sum_{\stackrel{g' \in N(f)}{g' \neq g}} \mu_{g' \rightarrow f}(\mathbf{Z_{g' \cap f}})\big]
$}
\label{eq:between-functions}
\end{eqnarray}
where $\mathbf{Z_{f \cap g}}$ stands for an assignment to the variables in the scope of $f$ and $g$, $\mathbf{Z_{f - g}}$ stands for an assignment to the variables in the scope of $f$ that are not in $g$, $N(f)$ stands for the set of functions liked to $f$ ($f$'s neighboring functions), and $\mu_{g' \rightarrow f}$ stands for the message from function $g'$ to $f$. 

Observe in Figure \ref{fig:ms-independent-binary} that selection and cost functions are connected by a single binary variable. Thus, the messages exchanged between functions in our problem will refer to the assignments of a single binary variable.  In other words, the assignment $\mathbf{Z_{f \cap g}}$ will correspond to some binary variable $z_{pr}$. Therefore, a message between functions must contain two values, one per assignment of a binary variable. At this point, we can make a further simplification and consider sending the difference between the two values. Intuitively, a function sending a message with a single value for a binary variable transmits the difference between the variable being active and inactive. In general, we will define the single-valued message exchanged between two functions as
\begin{eqnarray}
\nu_{f \rightarrow g} = \mu_{f \rightarrow g}(1) - \mu_{f \rightarrow g}(0)\ .
\label{eq:single-valued-message}
\end{eqnarray}
Next, 
we compute the messages between cost and selection functions.

\noindent\emph{(1) From cost function to selection function.} 
This message expresses the difference for a UAV $\rho_p$ between serving request $\tau_r$ or not, therefore
\begin{equation}
\nu_{v_{pr} \rightarrow s_{r}} = v_{pr}(1) - v_{pr}(0) = \delta_{pr} - 0 = \delta_{pr}\ .
\label{eq:cost-to-selection}
\end{equation}

\noindent
\emph{(2) From selection function to cost function.} Consider selection function $s_r$ and cost function $v_{pr}$.
From equation \ref{eq:between-functions}, we obtain
$$
\mu_{s_r \rightarrow v_{pr}}(1) = 0,
\ \mbox{ and }\
\mu_{s_r \rightarrow v_{pr}}(0) = \min_{\rho_{p'} \in P_r-\rho_p} \delta_{p'r} \ .
$$
Then we can apply equation \ref{eq:single-valued-message} to obtain the single-valued message $\nu_{s_r \rightarrow v_{pr}} = - \min_{\rho_{p'} \in P_r - \rho_p} \delta_{p'r}$. Moreover, this message can be computed efficiently. Consider the pair $\langle \nu^*,\nu^{**}\rangle$ as the two lowest values received by the selection function $s_r$. Then, the message that $s_r$ must send to each $v_{pr}$ is
\begin{eqnarray}
 \nu_{s_r \rightarrow v_{pr}} =
 \begin{cases}
 - \nu^* &  \nu_{v_{pr} \rightarrow s_r} \neq \nu^*\\
 - \nu^{**} & \nu_{v_{pr} \rightarrow s_r} = \nu^*
 \end{cases}.
\label{eq:selection-to-cost}
\end{eqnarray}

To summarize, each cost function computes and sends messages using equation \ref{eq:cost-to-selection}; each selection function computes and sends messages using equation \ref{eq:selection-to-cost}.

\noindent\textbf{Max-Sum operation}.
\label{sec:independent-operation}
Max-Sum is an approximate algorithm in the general case, but it is provably optimal on trees.
Due to how we encoded the problem, the resulting factor graph contains a disconnected, tree-shaped component for each request $r$ (see Figure \ref{fig:ms-independent-binary}). Thus, Max-Sum operates optimally in this case. Moreover, the algorithm is guaranteed to converge after traversing the tree from the leaves to the root and then back to the leaves again. In our case, the tree-shaped component for each request is actually a star-like tree, with the selection function $s_r$ at the center, and all others connected to it. Hence, we are guaranteed to compute the optimal solution in two steps if we pick $s_r$ as the root node of each component.

Typically, Max-Sum's decisions are made according to the costs seen by the variable nodes after running the algorithm. However, we have no variables in our graph anymore because we eliminated them. As a consequence, we have to make the decisions at either the selector nodes $s_r$ or at the cost nodes $v_{pr}$. The best option is to let the selectors choose, because it guarantees that the same task is never simultaneously assigned to two different UAVs. Because the decisions are made by the $s_r$ nodes, there is no need for the second Max-Sum iteration (messages from selector to cost functions) anymore.

Notice that, after our refinements, the only remaining Max-Sum nodes are the cost functions $v_{pr}$ and a selection function $s_r$ for each request. Hence, the UAVs can execute the algorithm distributedly by letting the cost functions $v_{pr}$ run on UAV $\rho_p$, and the selection functions $s_r$ run on the current owner of the request. Moreover, a UAV only needs to know which requests do its neighbors have to build the logical nodes it must run. Thus, the whole graph can also be built distributedly based only on the UAVs' local information.
 

Max-Sum runs on our example as follows. First, each leaf cost function $v_{pr}$  must send its cost to the root of its tree, $s_r$. That is, UAV $\rho_1$ sends $1$ to $s_3$ (within UAV $\rho_2$), and $5$ to $s_2$ (within itself). Likewise, UAV $\rho_2$ sends $2$ to $s_2$ and $2$ to $s_3$, whereas UAV $\rho_3$ sends $7$ to $s_1$. 
Thereafter, the $s_r$ nodes decide by choosing the UAV whose message had a lower cost. Hence, $s_3$ (running within UAV $\rho_2$) decides to allocate $\tau_3$ to $\rho_1$, $s_2$ allocates $\tau_2$ to $\rho_2$, and $s_1$ allocates $\tau_1$ to $\rho_3$. At this point the owner of each request knows whether a request should be reallocated or not and to whom, so we can use it to implement the reallocation phase described in Section~\ref{sec:approach}.


\section{Coordination using Workload-based Valuations}
\label{sec:workload}

In realistic scenarios, requests do not appear uniformly across time and space but 
concentrated around one or several particular areas, hereafter referred to as \emph{hot spots}. In that case, the assumption of independence in the valuation of the requests provides an allocation that  assigns a large number of requests  to the UAVs close to the hot spot, leaving the remaining UAVs
idle. Thus, the independence assumption is too strong because it leads to utterly unbalanced allocations.
Next, we show that it is possible to relax this assumption while keeping an acceptable time complexity for Max-Sum. We introduce a new factor for each UAV: a penalty that grows as the number of requests assigned to the UAV increases.  Formally, let $Z_p= \{z_{pr}|\tau_r \in R_p\}$ be the set of variables encoding the assignment to UAV $\rho_p$. The number of requests assigned to UAV $p$ is $\boldsymbol{\eta}_p = \sum_{r\in R_p}\mathbf{z}_{pr}$. The workload factor for UAV $\rho_p$ is
\begin{equation}
w_p(\mathbf{Z}_p) = f(\boldsymbol{\eta}_p) = k\cdot   (\boldsymbol{\eta}_p)^\alpha \quad ,
\label{eq:workload-factor}
\end{equation}
where $k\geq 0$ and $\alpha\geq 1$ are parameters that can be used to control the fairness in the distribution of requests (in terms of how many requests are assigned to each UAV). Thus, the larger the $\alpha$ and the $k$, the fairer the request distribution. 

The direct assessment of  Max-Sum messages going out of  the workload factor takes $O(N\cdot2^{N-1})$ time, where $N= |Z_p|$. Interestingly, the workload factor is a particular case of a \emph{cardinality potential} as defined by Tarlow \emph{et.} al.~\citep{Tarlow2010}. A cardinality potential is a factor defined over a set of binary variables ($Z_p$ in this case) that does only depend on the number of active variables. That is, it does not depend on which variables are active, but only on how many of them are active. As described in \citep{Tarlow2010}, the computation of the Max-Sum messages for these potentials can be done in  $O(N \log N)$. Thus, using Tarlow's result we can reduce the time to assess the messages for the workload factors from exponential in the number of variables to linearithmic.

To further speedup the algorithm and reduce message exchanges, we can add the workload factor and the cost factors that describe the cost for UAV $\rho_p$ to service each of the requests. 
Lemma~\ref{lem:IndepVal} in the appendix shows that if we have a procedure for determining the Max-Sum messages going out of a factor over binary variables, say $f$, we can reuse it to determine the messages going out of a factor $h$ that is the sum of $f$ with a set of independent costs, one for each variable.

\begin{algorithm}[!htb]
\caption{process$(v_p)$ \Comment{All undefined values are $\infty$}}
\label{alg:WorkloadFactor}
\begin{algorithmic}[1]
\begin{footnotesize}
    \For{$r \in R_p$} \Comment{Incorporate incoming messages}
        \State {$I[r] = \nu_{s_r \rightarrow v_p} + v_{pr}(1)$}
    \EndFor

    \State $Pos,I' = \mbox{sorted}(I)$  \Comment{$Pos$ are the reverse indices, and I' the sorted list}
    
    \State $cs = 0$
    \For{$i = 0 \mbox{ \textbf{to} } |I'|$} \Comment{Compute cumulative sums}
        \State $CS^0[i] = cs + w_p(i)$
        \State $CS^-[i] = cs + w_p(i-1)$
        \State $CS^+[i] = cs + w_p(i+1)$
        \State $cs = I'[i] + cs$
    \EndFor
    
    \For{$i = 0 \mbox{ \textbf{to} } |I'|$} \Comment{Compute cumulative mins}
        \State $M^+[i] = min(CS^+[i], M^+[i-1])$
        \State $M^L[i] = min(CS^0[i], M^L[i-1])$
        \State $M^R[i] = min(CS^0[|I'|-i], M^R[|I'|-i+1])$
        \State $M^-[i] = min(CS^-[|I'|-i], M^-[|I'|-i+1])$
    \EndFor
    
    \For{$r \in R_p$} \Comment{Compute and send messages}
        \State $i = Pos[r]$
        \State $\xi_0 = min(M^L[i-1], M^-[i+1] - I'[i])$
        \State $\xi_1 = min(M^+[i-1], M^R[i+1] - I'[i])$
        \State {$\nu_{w_p \rightarrow s_{r}} = \xi_1 - \xi_0 + v_{pr}(1)$}
        \State send$(\nu_{w_p \rightarrow s_{r}})$
    \EndFor
    \end{footnotesize}
\end{algorithmic} 
\end{algorithm}

Thus, we can define a single factor that expresses the complete cost of a UAV when assigned a set of requests, which amounts to the sum of the independent costs for each  of the requests assigned plus the workload cost for accepting that number of requests. Formally the cost factor for UAV $\rho_p$ is
\begin{equation}
w_p(\mathbf{Z}_p) + \sum_{\tau_r \in  R_p}v_{pr}(z_{pr}) \quad.
\label{eq:workload-evaluation}
\end{equation}

Algorithm~\ref{alg:WorkloadFactor} defines how to assess the messages flowing out of the cost factor $v_p$. It is obtained by composing Lemma~\ref{lem:IndepVal} with the algorithm described in \citep{Tarlow2010} for cardinality potentials.\footnote{The algorithm description provided in \citep{Tarlow2010} is inaccurate. However, the authors own implementation of the algorithm is correct. It does not match the description in \citep{Tarlow2010} and is closer to the description provided by Algorithm~\ref{alg:WorkloadFactor}.} The algorithm is basically a dynamic programming procedure for computing minimizations over cumulative sums, which can be done in a few linear passes. The complexity is dominated by the initial sort operation.

Summarizing, by introducing workload valuations that do not only depend on each individual request, but also on the number of requests,   we have shown that it is possible to relax the assumption of independence between valuations with a very minor impact on the computational effort required to assess the messages (from linear to linearithmic). 
In the next section we show that this relaxation provides significantly 
better allocations in terms of median service time.

\section{Evaluation methodology}
\label{sec:methodology}

In the following sections we want to empirically evaluate our MRF-based solutions for the LORP. The evaluation will be performed using the MASPlanes~\citep{MASPlanes} simulation environment. MASPlanes allows us to generate scenarios of varying characteristics, and then run any of the implemented algorithms against the exact same problem instances to see how they compare.

Comparing our algorithm's performance against the current state-of-the-art is tricky because, as explained in Section~\ref{sec:state-of-art}, most methods can not cope with the requirements of our problem domain. Therefore, we implemented a relaxed version of the problem to compare against them. In this relaxation, the allocation is performed by a central agent that has no communication restrictions. Furthermore, the centralized allocation procedure used by the centralized agent is considered to be instantaneous. That is, there is no delay at all introduced by the allocation procedure, whatever its computational cost may be. 
The only constraint that such central planner enforces is that a UAV can not be assigned to a request whose existence is unknown by that UAV. 
In the following experiments, the central agent employs one of two different request allocation algorithms to make decisions using this relaxed problem. 

First, the \emph{c-hungarian} algorithm runs the Hungarian method~\citep{kuhn1955hungarian} to compute an allocation. This method optimally solves the assignment problem in $O(n^3)$ time. However, recall that our problem is not an assignment problem because: (i) unassigned requests will have to be performed later on; (ii) the cost of servicing a request changes depending on the order; and (iii) new requests may be introduced at any future time. Also notice that the solutions obtained using this method correspond to those that would be found by the auction algorithm~\citep{bertsekas1988auction}. Aside from the actual procedure, the only difference between these methods is that the auction algorithm may be distributed, at the expense of additional delays caused by communications. Hence, the results that we obtain for this method provide an upper bound on the best results that a distributed auction algorithm would obtain.

\begin{algorithm}[tb]
\caption{GreedyAllocation$(P, R, V)$}
\label{alg:GreedyAllocation}
\begin{algorithmic}[1]
\small
\renewcommand{\algorithmicrequire}{\textbf{Input:}}
\Require {$P$ is a set of planes}
\renewcommand{\algorithmicrequire}{\textbf{\color{white}Input:}}
\Require {$R$ is a set of requests}
\Require {$V$ is a map where each $V[p]$ are all requests known by $p \in P$}
	\State $A[p] = \emptyset, \forall p \in P$
	
	\While{$R \neq \emptyset$}
		\State $\langle p^*, r^* \rangle = \arg\min_{p \in P, r \in R | r \in V[p]} \mbox{\emph{evaluate}}(r, A[p])$ 
	    \State $A[p^*]$ = insert$(r^*, A[p^*])$
	    \State $R = R \setminus \{r^*\}$
	\EndWhile
	\renewcommand{\algorithmicensure}{\textbf{Output:}}
\Ensure {$A$, a map where each $A[p]$ are all requests allocated to $p$}
\end{algorithmic} 
\end{algorithm}

Second, \emph{c-greedy} represents a straightforward sequential greedy algorithm as presented in Algorithm~\ref{alg:GreedyAllocation}. Initially, the central agent considers that no request is allocated to any UAV. Then, it computes the best allocation (the allocation with minimum cost) of a single task to a single UAV. Thereafter, it keeps finding the best possible allocation of a single request to an UAV, but taking into account the previously allocated requests. The process ends when all requests have been assigned to some UAV. This procedure can be seen as a centralized version of the Sequential Single Item~\citep{koenig2010progress} auctions mechanism. Hence, this solution represents an upper bound of the results we would be able to achieve while using SSI auctions if we were somehow able to use them. Because we want to minimize the average service time, the \emph{c-greedy} algorithm employs the recommended \emph{BidMinPath} bidding rule from~\citep{lagoudakis2005auction} as the \emph{evaluate()} function in Algorithm~\ref{alg:GreedyAllocation}. Finally, notice that this method is also equivalent to the \emph{SGA} method presented in~\citep{choi2009consensus}.
This implies that, as proven in that work, the decisions obtained by \emph{c-greedy} are equivalent to the solutions that would be obtained by the Consensus-Based Bundle Algorithm (CBBA) in the ideal case of a static network structure. Thus, \emph{c-greedy} also represents an upper bound on the quality of the solutions that CBBA would obtain if we were able to employ it.

	For completeness, we also implemented centralized versions of our algorithms: \cindependent, that represents our MRF-based solution using independent valuations; and \cworkload, that uses workload valuations as detailed in Section~\ref{sec:workload}. These work exactly like the distributed versions, but within a single central agent and thus without latency penalties when messages are exchanged.

All these centralized algorithms serve as baselines to assess the performance of the distributed solutions. Obviously, the distributed solutions include our proposed solutions \dindependent (using independent valuations) and \dworkload (using workload valuations). Additionally, we implemented the distributed state-of-the-art PSI method, and obtained very interesting results: the PSI method and our \dindependent method provide exactly the same decisions. This can be seen by following the example execution in Section~\ref{sec:independent-operation}, and interpreting it as if request owners opened an auction for each of the tasks they have, and neighboring UAVs bid for them (lowest bid wins). Due to this exact functional match between PSI and our solution, we only report results about \cindependent and \dindependent, but these represent also the results obtained by the PSI method.
At this point, a keen reader may ask whether we could employ workload evaluations (from eq. \ref{eq:workload-evaluation}) in the PSI method. However, recall that in PSI all tasks are auctioned in parallel. Thus, when two UAVs meet and one has several more tasks than the other, the UAV with fewer tasks will actually win \emph{all} tasks (at the same time) because it has a much lower workload. In the next reallocation cycle, the entirely opposite happens, and so on. In the end, the result is a thrashing behavior that blocks both planes from performing any useful work.

\begin{table}[t]
\centering
\begin{tabular}{|l | l l|}
\hline
\textbf{Method} 		& \textbf{is the ideal implementation of} & \textbf{if disregarding} \\
\hline\hline 
c-hungarian			& Auction algorithm	& CD, SA\\
\hline \multirow{2}{*}
{c-greedy}			& CBBA				& CD, ND, SA\\
					& SSI auctions		& CD, ND, SA\\
\hline \multirow{2}{*}
{c-independent}		& PSI auctions 		& CD, SA \\
					& Independent valuations & CD, SA\\
\hline
c-workload			& Workload valuations & CD, SA\\
\hline \multirow{2}{*}
{d-independent}		& PSI auctions 		& - \\
					& Independent valuations & -\\
\hline 
d-workload			& Workload valuations & -\\
\hline
\end{tabular}
\caption{Relationship of the evaluated algorithms with other state-of-the-art approaches. CD, DN and SA stand for Communication Delays, Network Dynamicity and Situational Awareness respectively.
\label{tab:method-equivalences}}
\end{table}


To Summarize, Table~\ref{tab:method-equivalences} presents the algorithms we test and their relationship with other state-of-the-art approaches. Under the assumptions presented in that table, the algorithm's decisions would match those that we obtain in our evaluation. However, this does not necessary imply that the algorithms can not work without such assumptions. For instance, CBBA can operate on a dynamic network, but would obtain worse results than those of \emph{c-greedy}.


\section{Effects of the spatial distribution of requests}
\label{sec:results}

In this section we aim to validate the hypothesis that different spatial distributions of the incoming requests may favor some methods over others. Because this assumption is independent of whether the algorithms can or cannot work in a distributed manner, we start by comparing all algorithms using their idealistic (centralized) implementation. Thereafter, we compare the results of \cindependent and \cworkload with their actually applicable \dindependent and \dworkload distributed counterparts.

\subsection{Empirical settings}

We prepared two scenarios that represent a time-span ($T$) of one month. During that time, 10 UAVs with a communication range of \unit[2]{km} survey a square field of \unit[100]{km$^2$}. We assume that the UAVs always travel at a cruise speed of \unit[50]{km/h}. In these scenarios, a single operator submits requests at a mean rate of one request per minute. However, we introduce four crisis periods during which the rate of requests is much higher. To accomplish this, the requests submission times are sampled from a mixture of distributions. The mixture contains four normal distributions $\mathcal{N}_i(\mu_i, \unit[7.2]{h})$ (one per crisis period) and a uniform distribution for the non-crisis period. The $\mu_i$ means themselves are sampled from a uniform distribution $\mathcal{U}(T)$.

\begin{figure}[b]
	\centering
	\begin{minipage}[b]{.42\linewidth}
		\centering
		\raisebox{-0.5\height}{\includegraphics[width=\linewidth]{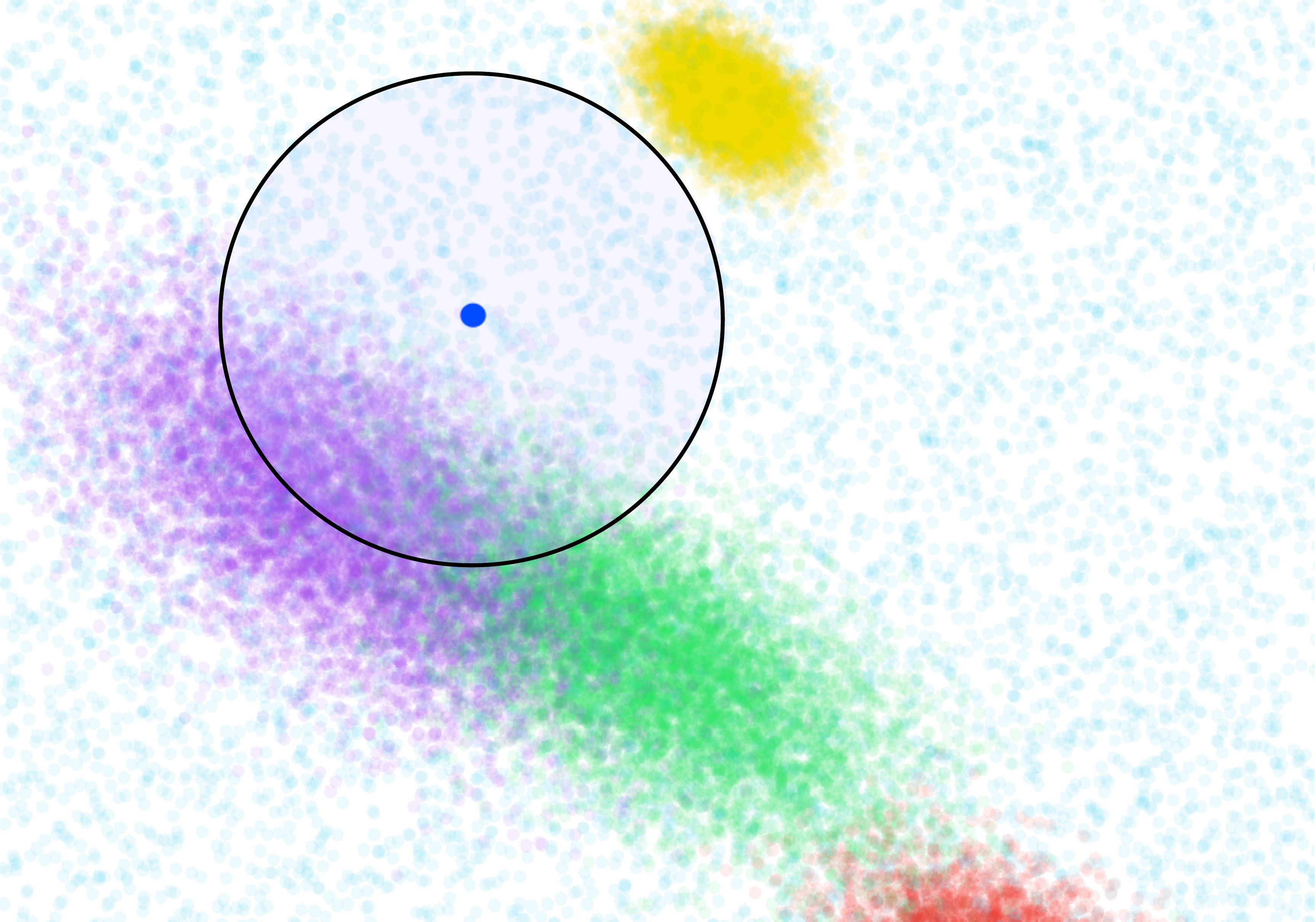}}
		\caption{Example \emph{hot spots} requests distribution.}
		\label{fig:gaussian_scenario}
	\end{minipage}
	\hfill
	\begin{minipage}[b]{.55\linewidth}
		\centering
		\includegraphics[width=\linewidth]{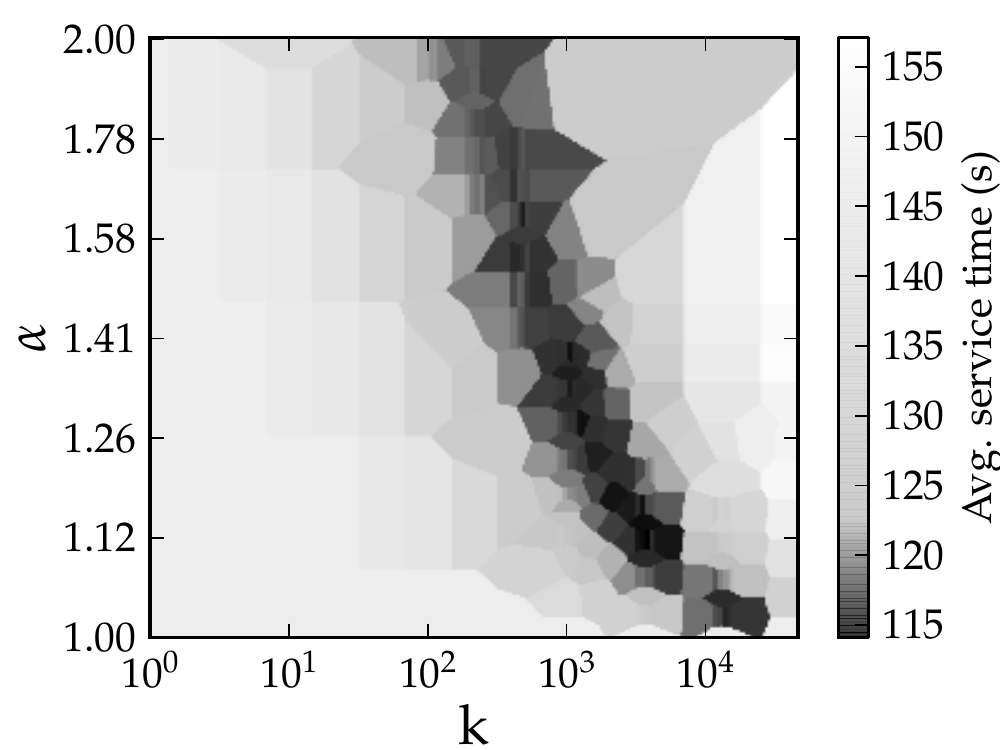}
		\caption{Parameter exploration in the \emph{hot spots} scenarios.}
		\label{fig:parameters}
	\end{minipage}
\end{figure}


The two scenarios differ on the spatial distribution of requests. In the \emph{uniform} scenario, requests are uniformly distributed along space, whereas the \emph{hot spot} scenario models a more realistic setting where crisis requests are localized around hot spots. These spatial hot spots are defined as bivariate Gaussian distributions with randomly generated parameters. Figure~\ref{fig:gaussian_scenario} depicts an example of such scenario, where colored dots stand for requests. The scattered dots correspond to non-related requests, whereas related requests form dot clouds around their hot spot. Finally, the strong dot represents the operator, and the circle surrounding it represents its communication range.


To use the \emph{c-workload} method we have to set values for the $k$ and $\alpha$ parameters. Hence, we performed an exploration on the space of these parameters to determine which values are suitable to the \emph{hot spot} scenarios. Figure~\ref{fig:parameters} shows the results we obtained after this exploration. The colors correspond to the median of the average service time that we obtained after running the algorithm in 30 different scenarios for each pair $(k,\alpha)$. For instance, when $k=10^2$ and $\alpha=1.12$ the algorithm achieved a median average service time of \unit[137]{s}.
Observe that the algorithm exhibits a smooth gradient for any fixed value of $\alpha$ or $k$. Hence, good combinations of $k$ and $\alpha$ can be found by fixing one parameter to a reasonable value and performing a descent search on the other one. We chose $k=1000$, and found the best corresponding $\alpha$ to be $1.36$ with $0.01$ precision. We also conducted a similar exploration in the uniform scenario. However, the results were similar for $k$ values between $10^2$ and $10^5$, and $\alpha$ values between $1.01$ and $1.68$. Hence, we use the same values than in the hot spot scenario.

\subsection{Results}

\begin{figure}
	\centering
	\includegraphics[width=1\linewidth]{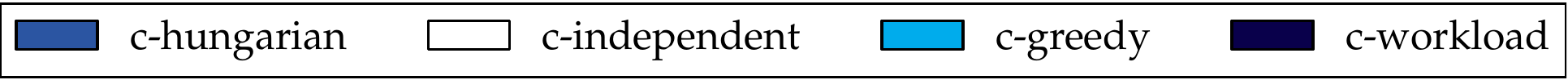} \\
	\begin{subfigure}{.49\linewidth}
		\centering
		\includegraphics[width=\linewidth]{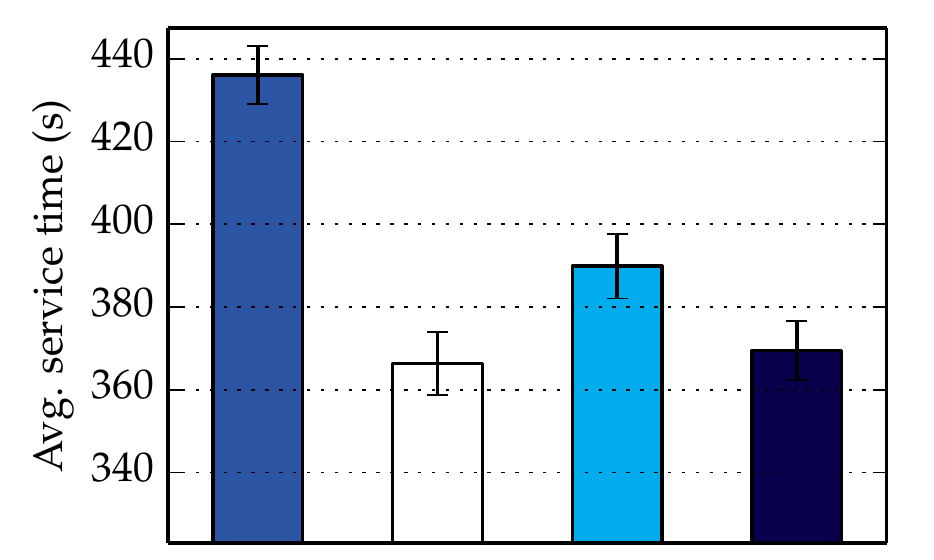}
		\caption{\emph{uniform} scenario}
		\label{fig:results-centralized-uniform}
	\end{subfigure}
	\begin{subfigure}{.49\linewidth}
		\centering
		\includegraphics[width=\linewidth]{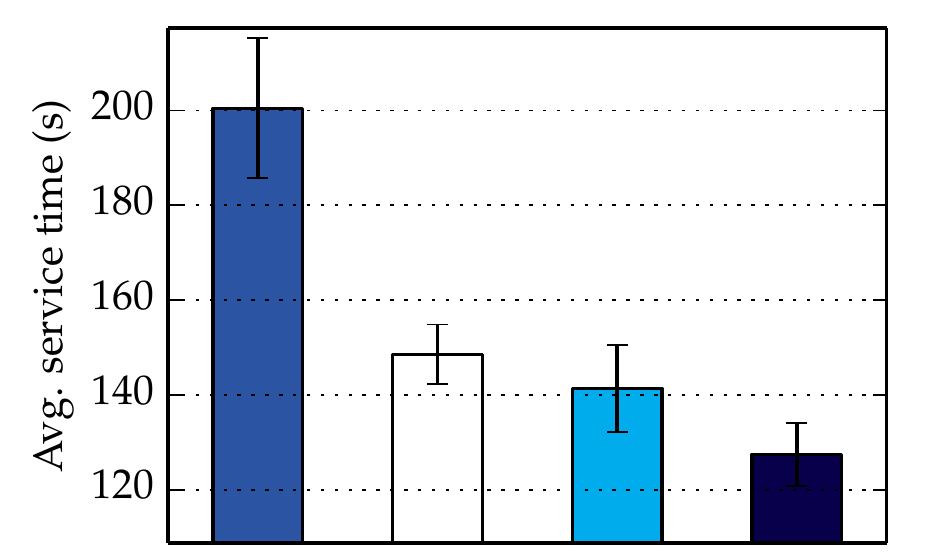}
		\caption{\emph{hot spots} scenario}
		\label{fig:results-centralized-gaussian}
	\end{subfigure}
	\caption{Effects of the spatial distribution of requests. Error bars show the standard error of the mean.}
	\label{fig:results-spatial-distribution}
\end{figure}

Next we ran all centralized algorithms on a new set of 30 new problems of each scenario to ensure that the parameters were not overfitted. Figure~\ref{fig:results-centralized-uniform} shows the average service time (time passed between when the request is generated and when the planes service it) achieved by the different algorithms in the \emph{uniform} scenario. The best performing algorithm is \cindependent, with \cworkload being just 1\% worse. Despite the small difference, all results are found significant by the (paired, nonparametric) Wilcoxon signed rank test~\citep{wilcoxon1945individual} with $p=0.01$. The worst algorithm is clearly \chungarian, taking nearly 16\% more time on average.

These results contrast with those of the \emph{hot spot} scenario. As shown in Figure~\ref{fig:results-centralized-gaussian}, \cworkload rises as the best method in this case, taking 10\% less time than \cssi and 14\% less time than \cindependent. Interestingly, this means that both \cworkload and \cssi are able to capture and exploit the underlying spatial distribution of requests, whereas \cindependent works very well when the requests are randomly distributed but is not well equipped to handle other, more realistic distributions. Finally, \chungarian's results show that completely disregarding the dynamism of the problem leads to overall bad decisions.

\begin{figure}
	\includegraphics[width=1\linewidth]{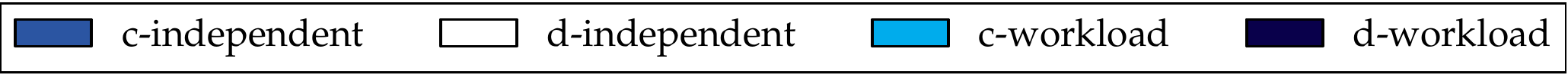} \\
	\begin{subfigure}{.49\linewidth}
		\centering
		\includegraphics[width=\linewidth]{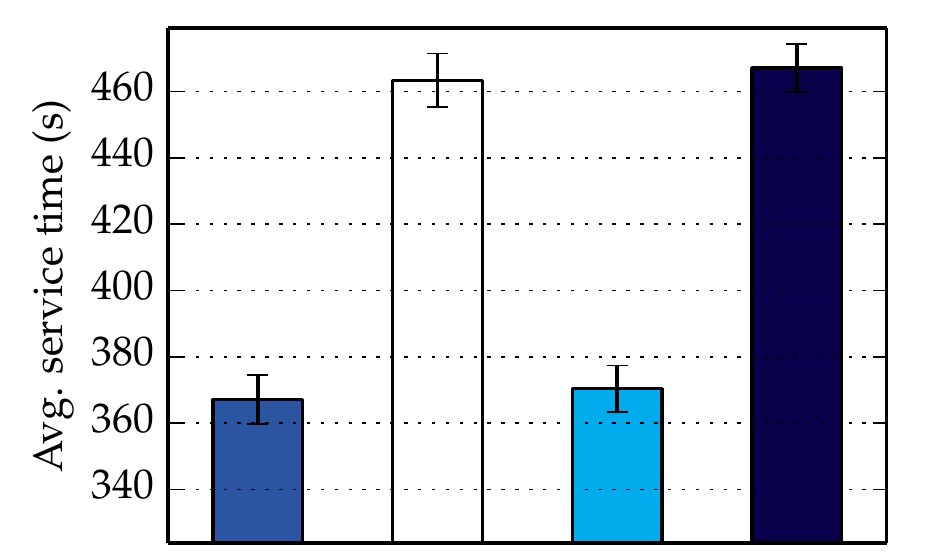}
		\caption{\emph{uniform} scenario}
		\label{fig:results-uniform}
	\end{subfigure}
	\hfill
	\begin{subfigure}{.49\linewidth}
		\centering
		\includegraphics[width=\linewidth]{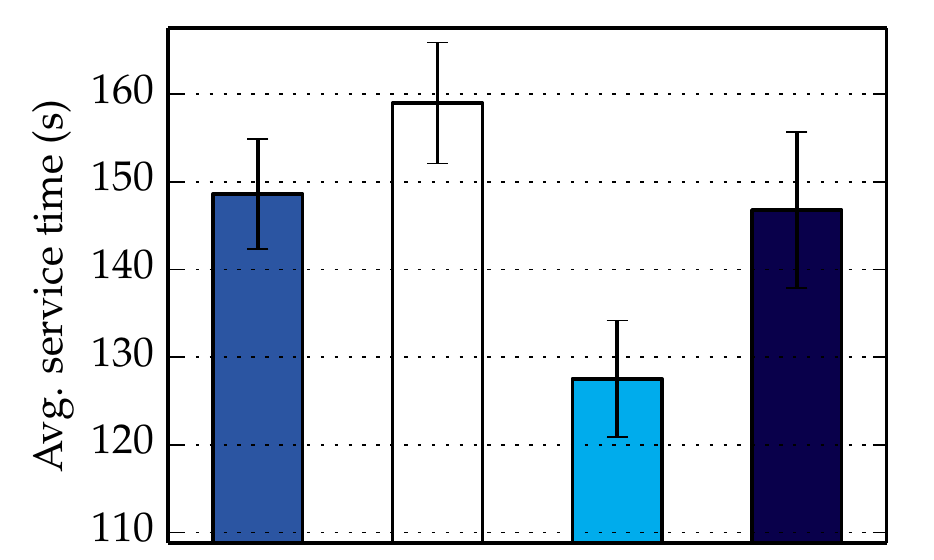}
		\caption{\emph{hot spots} scenario}
		\label{fig:results-gaussian}
	\end{subfigure}
	\caption{Impact of distributed solving.}
	\label{fig:results-cost-distribution}
\end{figure}

At this point, it is clear that the underlying spatial distribution of tasks has different effects on the varying methods. However, we must check whether such differences still hold in the actually applicable distributed versions. Hence, we ran the same scenarios, but this time using \dindependent and \dworkload to compare their results with the respective idealized cases. Figure~\ref{fig:results-cost-distribution} shows the results obtained on both scenarios. Notice that the differences persist, with \dindependent achieving insignificantly ($p=0.01$) better results than \dworkload in the uniform scenario while being significantly worse in the \emph{hot spots} one. In fact, \dworkload turns out to be as good as \cindependent in that scenario according to the Wilcoxon test.

Aside from confirming our hypothesis about the spatial distribution of requests, this experiment also provides an idea about the cost of distributed solving. Specifically, it shows that there's a 20\% time increase between the distributed methods and their idealized counterparts in the \emph{uniform} scenario, and between 6\% (\{c,d\}-independent) and 13\% (\{c,d\}-workload) in the \emph{hot spots} one. The lower overhead in the \emph{hot spots} scenario is due to requests being closer between them, giving the planes more opportunities to coordinate.

\section{Exploring d-workload's behavior}
\label{sec:results2}

In the previous Section we presented results on two scenarios that differed only on the spatial distribution of requests. However, it is still unclear how \dworkload will perform in other scenarios. Hence, in this section we aim at providing a better inside on the behavior of \dworkload under varying scenario's characteristics. 

\subsection{Empirical settings}

To this end, we chose a number of scenario's characteristics and prepared a full factorial experiment to evaluate their impact on the different algorithms. To capture a wide range of scenarios yet keep the number of experiments within a reasonable number, we selected the following four characteristics:

\begin{itemize}\item \textbf{Load.} Defines the overall amount of requests per UAV. To keep things simple, we always introduce a total of $43,200$ requests (\unit[1]{req/min} on average), so the load is controlled solely by varying the number of available UAVs.
\item \textbf{Spread.} Captures how spatially spread the requests are between them. In scenarios with low spread, the requests tend to be clustered together around hotspots. In high-spread scenarios, the tasks are scattered around a larger area. We control the spread of the scenario by defining a \emph{hostspot-radius}. In the problems of that scenario, approximately 90\% of the tasks of each hotspot will be within hotspot-radius distance from the hotspot center. Such distribution is achieved by sampling from an Inverse Wishart distribution with $2.5$ degrees of freedom and a scale matrix experimentally adjusted to produce locations within the desired hotspot radius.
\item \textbf{Communication Range.} Represents the UAVs' (and operator) communication ranges.
\item \textbf{Time distribution Sharpness.} Specifies how constant is the rate at which requests are introduced. Because we keep a constant number of tasks for all scenarios, sharpness can be defined fully in terms of the number of crises. In high sharpness scenarios, there is a single short and highly active crisis period. In contrast, low sharpness scenarios contain nine different crisis periods of moderate activity, but no highly active period.
\end{itemize}

The chosen values for each feature are detailed in Table~\ref{tab:features}. For the characteristics not described here we employed the values reported in Section~\ref{sec:results}. Namely, the scenarios represent a full month of simulated time, with planes that move at a constant speed of \unit[52]{km/h} on a square area of \unit[100]{$\mbox{km}^2$}.

\begin{table}[tb]
\centering
\begin{tabular}{|lll|}
\hline
\textbf{Feature} & \textbf{Parameter} & \textbf{(Low, Mid, High)} \\
\hline \hline
Load & number of planes & 20, 10, 5 \\ \hline
Spread & hotspot-radius & \unit[1]{km}, \unit[3]{km}, \unit[6]{km} \\ \hline
Communication Range & communication range & \unit[1]{km}, \unit[2]{km}, \unit[3]{km} \\ \hline
Time distribution Sharpness & number of crises & 9, 3, 1 \\
\hline
\end{tabular}
\caption{Problem dimensions explored and their values}
\label{tab:features}
\end{table}

\begin{figure}
\centering
\includegraphics[width=.7\linewidth]{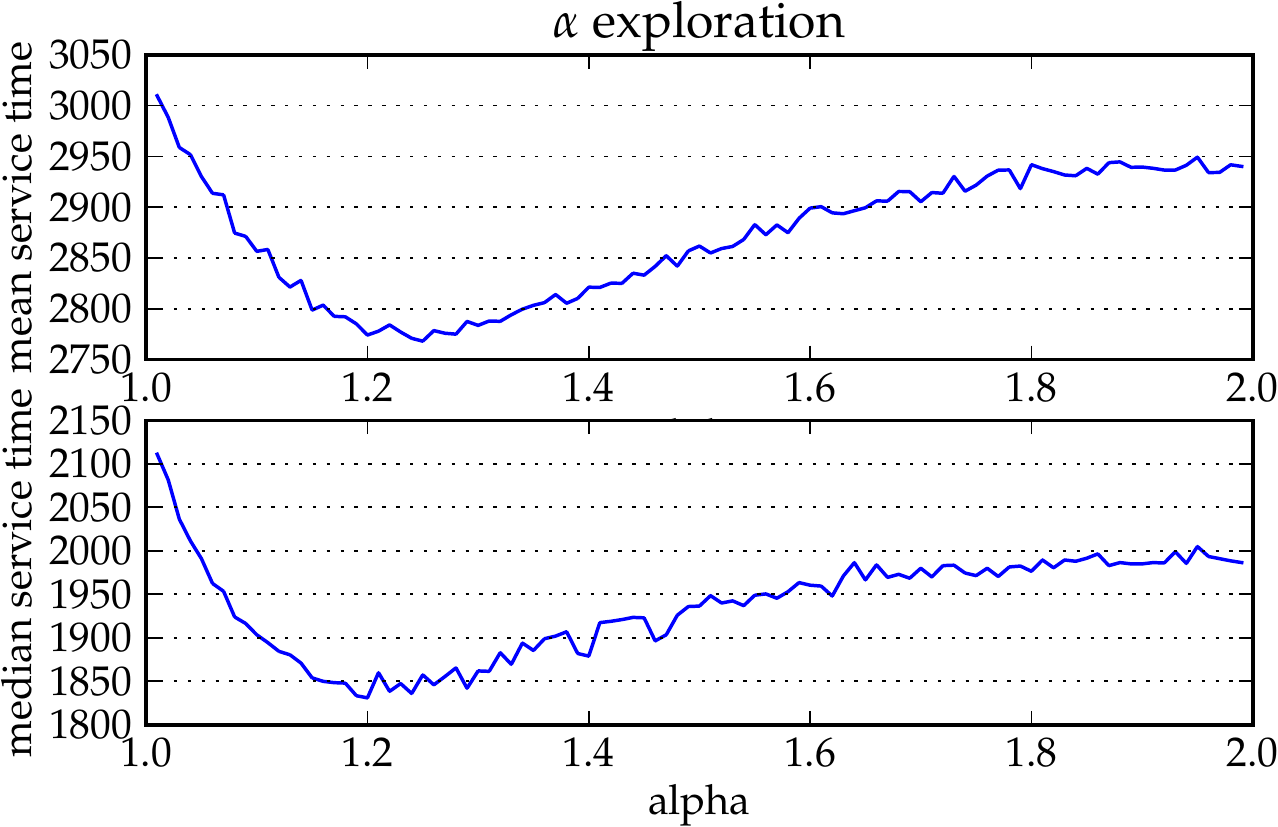} 
\caption{Exploration of $\alpha$ values.}
\label{fig:alpha-exploration}
\end{figure}

Once again, we must decide the $k$ and $\alpha$ values to use in our \dworkload method. From the previous results, we know that a good value for $\alpha$ can be found for any reasonable $k$ value. Therefore, we fixed $k=1000$ and explored on the $\alpha$ parameter. Thus, we generated an example problem from each of the above 81 scenarios, and ran \emph{d-workload} with $k=1000$ and $\alpha=[1.01, 1.02, .., 2]$ on each of them. Figure~\ref{fig:alpha-exploration} shows the mean and median average request service times obtained for each $\alpha$ value. In those graphs we observe a (somewhat rough) gradient, as expected by the previous results. Overall, the best $\alpha$ value is $\alpha=1.25$, with a median service time of \unit[276.9]{s} and a mean service time of \unit[183.1]{s}.

\subsection{Results}

After choosing the $k$ and $\alpha$ values, we are ready to compare all algorithms in the different scenarios. With this aim, we generated 30 problems of each scenario (for a total of 2430 problems). On each of these problems we evaluate the performance of: (i) \dindependent, which represents both the state-of-the-art PSI auctions method as well as our MRF-based solution with independent valuations; (ii) \cssi, that came up as the best existing centralized method; (iii) \dworkload, which is our novel proposed algorithm; and (iv) \cworkload, which represents a bound on the best performance achievable by our novel \dworkload mechanism.

\begin{figure}
\centering
	\includegraphics[width=1\linewidth]{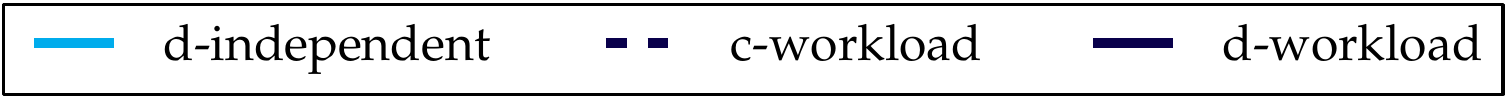}
	\\
	\begin{subfigure}{.49\linewidth}
		\includegraphics[width=\linewidth]{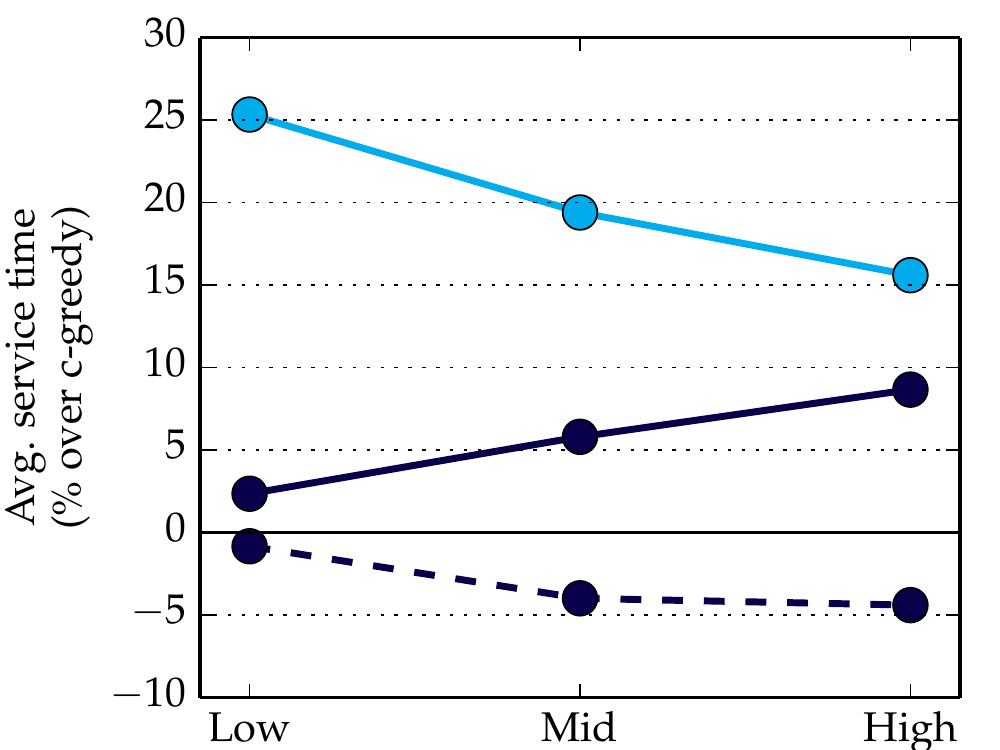}
		\caption{Load}
		\label{fig:main-effects-load}
	\end{subfigure}
	\hfill
	\begin{subfigure}{.49\linewidth}
		\includegraphics[width=\linewidth]{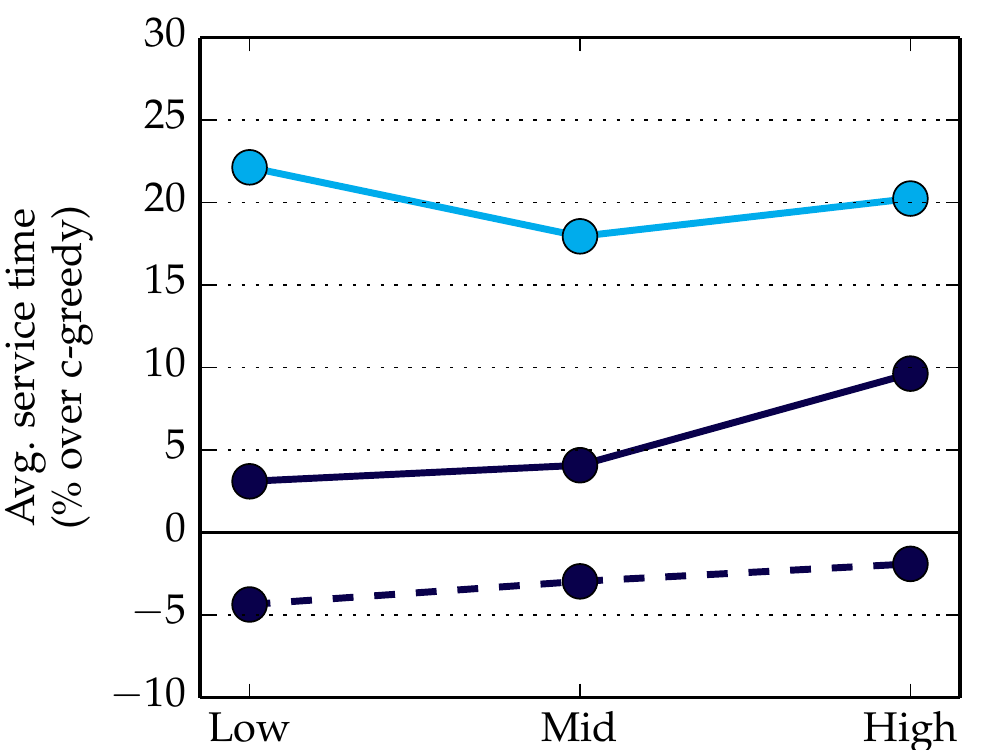}
		\caption{Spread}
		\label{fig:main-effects-spread}
	\end{subfigure}
	\\
	\begin{subfigure}{.49\linewidth}
		\includegraphics[width=\linewidth]{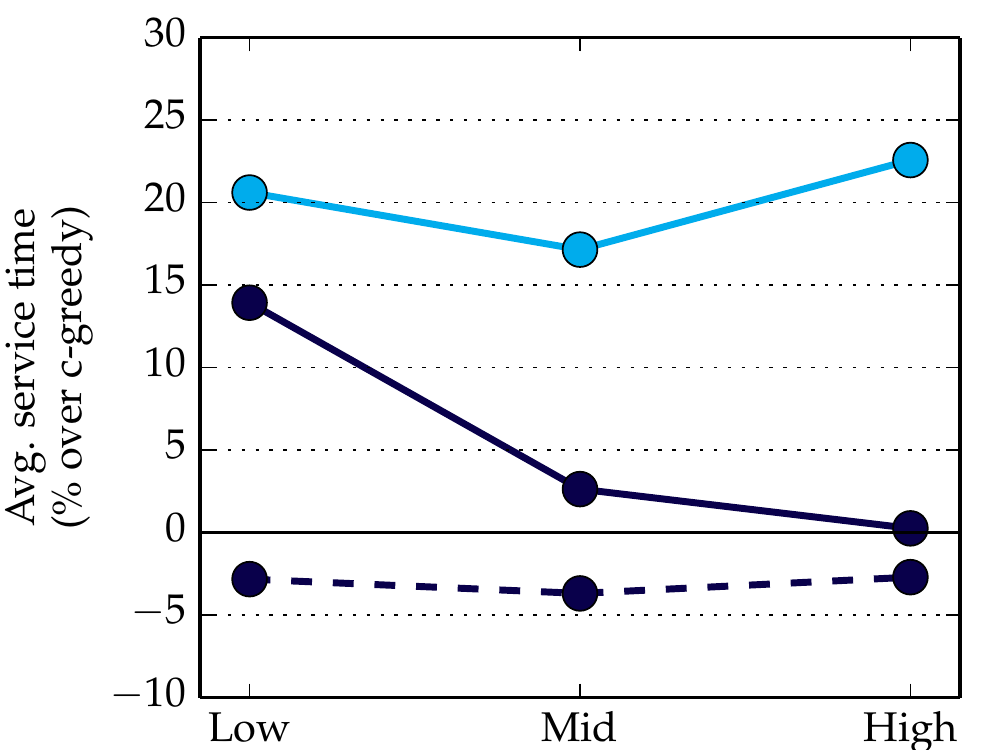}
		\caption{Communication Range}
		\label{fig:main-effects-range}
	\end{subfigure}
	\hfill
	\begin{subfigure}{.49\linewidth}
		\includegraphics[width=\linewidth]{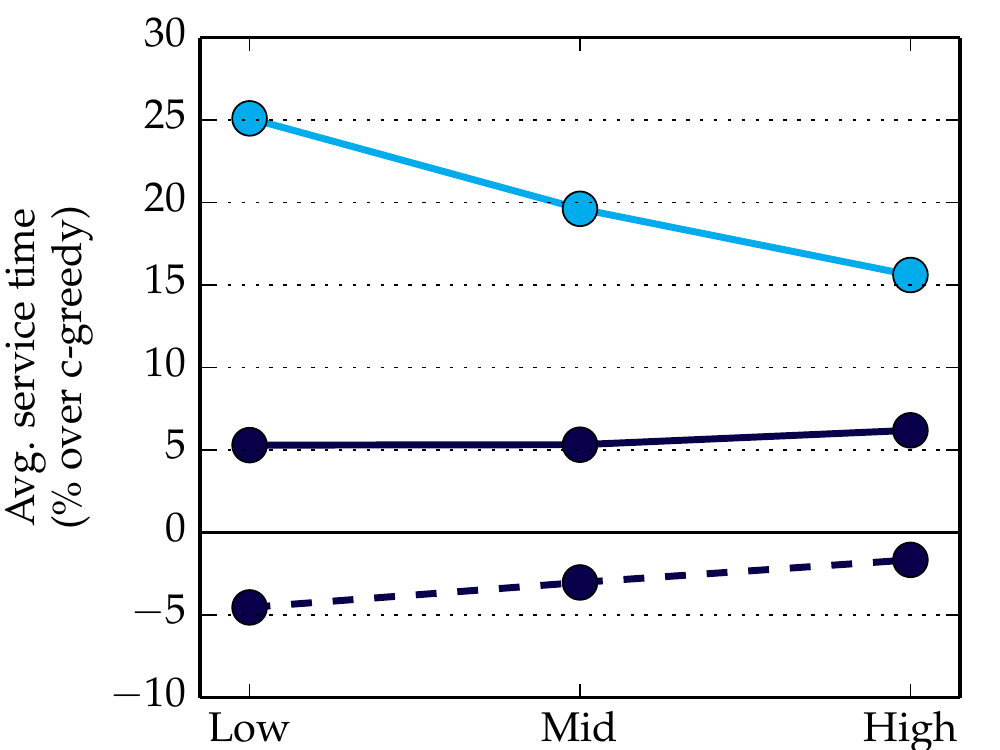}
		\caption{Time distribution Sharpness}
		\label{fig:main-effects-peakness}
	\end{subfigure}
\caption{Main effects plots (percentage \emph{w.r.t.} \cssi).}
\label{fig:main-effects}
\end{figure}

Figure \ref{fig:main-effects} shows the main effects plots of each characteristic, with the results of each algorithm represented as a percentage of the average service time with respect to that of \cssi. Albeit sometimes small, the differences between all methods are deemed significant by the Wilcoxon signed-rank test with $p=0.01$. Each plot shows the effect of the studied characteristic averaging across the levels of all other characteristics. For instance, the \emph{Low} point for \dworkload in Figure~\ref{fig:main-effects-load} represents the average service time of \dworkload between all problems with a \emph{Low} load, whatever their spread, communication range and sharpness levels are.

Figure~\ref{fig:main-effects-load} shows that the higher the load, the worse \dworkload compares to \cssi. However, notice that \cworkload displays an opposite behavior, becoming better than \cssi when there are more tasks per plane. This may seem counterintuitive at first, but has an easy explanation: recall that we control the load by varying the number of available planes. Hence, the higher the load, the less planes involved in the experiment. As a consequence, \dworkload's performance degrades not because there are more tasks, but because the planes are more spread and have less opportunities to coordinate. Moreover, \emph{d-workload} approaches \cssi's results when the number of planes increases (reading the graph from right to left), almost catching up to it despite operating in a distributed manner.

Figure~\ref{fig:main-effects-spread} shows that, unsurprisingly, \dworkload achieves comparatively worse results as the spatial distribution of requests becomes sparser. This is due to two reasons. First, because the more spread request are, the more UAVs will fly apart, and the lower chances to coordinate because of the communication range limit. Second, because workload methods exploit the spatial correlation between requests: if requests are more spread, this means that there is a lower spatial correlation between them, and hence the workload heuristic becomes worse. Finally, the previous section showed that \dindependent works comparatively better when the requests are uniformly distributed. This works at its favor when the spread increases. However, being distributed mechanism it is also negatively affected by the first reason above. As a consequence, the relative performance of \dindependent varies unpredictably when the spread changes.

In Figure~\ref{fig:main-effects-range} we can observe that, as the communication range increases, the distributed methods' results approach those of the centralized ones. This is because, as the communication range grows, the UAVs have more opportunities to coordinate. There is an exception in the case of \dindependent and a high communication range. In this case, a larger communication range actually worsens the results, because it causes the planes to stay very spread out, thus dealing much worse especially in the low spread scenarios. In contrast, \dworkload ends up achieving very close (less than 1\% worse) results to those of \cssi in the largest range we tested. 

Finally, Figure~\ref{fig:main-effects-peakness} also shows some interesting behaviors. On the one hand, \dindependent becomes better as the sharpness increases. This is because UAVs that coordinate with this method tend to evenly split the covered space between them. Hence, the more requests there are at the same time, the better its outcome. In contrast, \cssi and \dworkload try to be more clever than that, which pays off better when there are less requests than when the system is overloaded. This is even more noticeable with the workload-based methods. Recall that the main strength of such methods is to prepare for future requests. However, high sharpness scenarios reward better plans for the current requests than better predictions about the future.

Given that the only state-of-the-art method that satisfies all our problem's constrains is \dindependent, these graphs show a very interesting figure: \dworkload outperforms \dindependent on all scenarios, lowering the average service time between \unit[36]{s} (16\%) and \unit[18]{s} (6\%).
Hence \dworkload becomes the method of choice for the limited range online routing problem. Moreover, in our experiments \emph{d-workload} closes between 25\% and 100\% of the gap existing between \dworkload and the state-of-the-art centralized \cssi algorithm.
Finally, these results imply that, in the face of highly dynamic situations, it becomes more important to properly capture the distribution of incoming requests (as \dworkload does) than to compute better plans for the currently known ones (as \cssi does).


\section{Conclusions and future work}
\label{sec:conclusions}

This paper introduced the limited-range online routing problem, which requires that UAVs coordinate to serve requests submitted by external operators. To tackle this problem, we employed an MRF-based solution instead of the more common market-based approaches. Using a novel encoding of the problem and the max-sum algorithm, we showed that this approach can functionally mimic the operation of a decentralized parallel single-auctions approach.
While the re-discovery of a method on top of a different theoretical foundation may seem irrelevant, it is a significant result in our case. Namely, the MRF-based approach provides an easily extensible framework where the PSI method does not. Particularly, in this work we have shown that it is possible to introduce new factors to represent the workload of each UAV while maintaining low computational and communication requirements. Empirical evaluation shows that the improved version achieves up to $16\%$ lower service times than the single-auctions approach. Moreover, the actual performance comes very close to that of employing state-of-the-art centralized SSI auctions. Because of the communication range limit, centralized SSI auctions can not be implemented in the real-world. Therefore, our workload-based mechanism stands as a good choice for robust decentralized coordination with a communication range limit.

In the future, we would like to improve our approach and allocation methods 
in two ways. Firstly, we would like to allow for planes to notify the operators when their tasks are completed. This could be achieved by introducing new tasks into the system that represent such notifications. Also, some further signaling would be required to avoid losing tasks when a UAV fails (and hence disappears along with those tasks it is carrying). Secondly, we envision better cost functions for the UAVs. A clear improvement in that front would be to better consider the actual paths UAVs will follow while completing tasks~\citep{niendorf2016traveling}. Likewise, it would be desirable to introduce additional factors in the cost function to better control the spread of UAVs throughout space, as well as the need for UAVs to visit operators regularly to allow them to introduce new tasks in the system.


\section*{Acknowledgements}
Work partially funded by projects Collectiveware (TIN2015-66863-C2-1-R) and RASO (TIN2015-71799-C2-1-P).

\section*{References}

\bibliography{planes}

\section*{Appendix}


\appendix
\begin{lem}
\label{lem:AddIndependentVal}
Let $f(x_1,\ldots,x_n)$ be a factor over binary variables $x_1,\ldots,x_n$. Let $g(\mathbf{x}_i) = \gamma_i  \mathbf{x}_i$ be another factor defined only over variable $x_i$. Let $h(x_1,\ldots,x_n) = f(x_1,\ldots,x_n) + g(x_i)$ be the factor obtained by adding factors $f$ and $g$. 
Let 

\vspace{-1.2em}
{\small\begin{equation}
\label{eq:MessageDefinition}
 \mu_{f \rightarrow x_j}(\mathbf{x}_j,\nu_1,\ldots,\nu_n) = 
\min_{\mathbf{X_{-j}}} \left(f(\mathbf{X}) + \sum_{ k\neq j}\nu_k \mathbf{x}_k\right)
\end{equation}}
and 
{\small\begin{align}
\label{eq:NuDefinition}
 \nu_{f \rightarrow x_j}(\nu_1,\ldots,\nu_n) = \notag \\ = \mu_{f \rightarrow x_j}(1,\nu_1,\ldots,\nu_n) - \mu_{f \rightarrow x_j}(0,\nu_1,\ldots,\nu_n)
\end{align}}
 Then we have that:
{\small\begin{align}
\label{eq:FactorAddition}
 \nu_{h \rightarrow x_j}(\nu_1,\ldots,\nu_n) \notag = \\ = \begin{cases}
                                          \nu_{f \rightarrow x_j}(\nu_1,...,\nu_i+\gamma_i,\ldots,\nu_n)&j\neq i\\
                                          \nu_{f \rightarrow x_j}(\nu_1,\ldots,\nu_i+\gamma_i,\ldots,\nu_n) + \gamma_i&j=i\\
                                         \end{cases}
 \end{align}}
\end{lem}

\begin{pf}
see the suplementary material.
\end{pf}

In terms of the Max-Sum algorithm, this means that if we can efficiently compute the messages flowing out of $f$, we can also efficiently compute the messages flowing out of $h$.

\begin{lem}
\label{lem:IndepVal}
Let $f$ be a factor over binary variables $Y=\{y_1,\ldots,y_n\}$. Let $g(\mathbf{Y}) = \sum_{i=1}^n \gamma_i \mathbf{y}_i$ be another factor defined as the addition of a set of $n$ independent factors, one over each variable $y_i$. Let $h(\mathbf{Y}) = f(\mathbf{Y}) + g(\mathbf{Y})$ be the factor obtained by adding $f$ and $g$. Let

\vspace{-1.2em}
{\small\begin{align*}
	\mu_{f \rightarrow y_j}(\mathbf{y}_j, \nu_1, \ldots, \nu_n) = 
	\min_{\mathbf{Y}_{-j}} \left[ f(\mathbf{Y}) + \sum_{ k\neq j}\nu_k \mathbf{y}_k \right] , 
\end{align*}}
and
{\small\begin{align*}
\nu_{f \rightarrow y_j}(\nu_1,\ldots,\nu_n) = \\= \mu_{f \rightarrow y_j}(1,\nu_1,\ldots,\nu_n) - \mu_{f \rightarrow y_j}(0,\nu_1,\ldots,\nu_n).
\end{align*}}
We have that
{\small\begin{align*}
\nu_{h \rightarrow y_j}(\nu_1,\ldots,\nu_n) = \nu_{f \rightarrow y_j}(\nu_1+\gamma_1,\ldots,\nu_n+\gamma_n) + \gamma_j .
\end{align*}}
\end{lem}

\begin{pf}
see the supplementary material.
\end{pf}

%

\end{document}